\documentclass[12pt,preprint]{aastex}
\usepackage{emulateapj5,apjfonts}
\usepackage[obeyspaces]{url}
\usepackage{graphics}
\usepackage{amssymb}
\usepackage{threeparttable}
\usepackage{bbm}
\usepackage{amsmath,textcomp,array}
\usepackage{onecolfloat}
\usepackage{bm}

\lefthead{Heitmann et al.}
\righthead{The Last Journey. I.}

\begin{document}

\def\head{

\title{The Last Journey. I. An Extreme-Scale Simulation on the Mira Supercomputer}
\author{Katrin~Heitmann\altaffilmark{1}, Nicholas Frontiere\altaffilmark{1,2}, Esteban Rangel\altaffilmark{1,3},
Patricia Larsen\altaffilmark{1}, Adrian Pope\altaffilmark{2},
Imran Sultan\altaffilmark{1}, Thomas
Uram\altaffilmark{3}, Salman~Habib\altaffilmark{1,2}, 
Hal Finkel\altaffilmark{3},   
Danila Korytov\altaffilmark{1,4}, Eve Kovacs\altaffilmark{1}, Silvio Rizzi\altaffilmark{3}, Joe Insley\altaffilmark{3}, and Janet Y.K. Knowles\altaffilmark{3}}

\affil{$^1$ High Energy Physics Division,  Argonne National Laboratory, Lemont, IL 60439}

\affil{$^2$ Computational Science Division,  Argonne National Laboratory, Lemont, IL 60439}

\affil{$^3$ Argonne Leadership Computing Facility,  Argonne National Laboratory, Lemont, IL 60439}

\affil{$^4$ Department of Physics, University of Chicago, Chicago, IL 60637}

\date{today}

\begin{abstract}

The Last Journey is a large-volume, gravity-only, cosmological N-body simulation evolving more than 1.24 trillion particles in a periodic box with a side length of 5.025Gpc. It was implemented using the HACC simulation and analysis framework on the BG/Q system Mira. The cosmological parameters are chosen to be consistent with the results from the Planck satellite. A range of analysis tools have been run in situ to enable a diverse set of science projects and, at the same time, to keep the resulting data amount manageable. Analysis outputs have been generated starting at redshift $z\sim 10$ to allow for construction of synthetic galaxy catalogs using a semi-analytic modeling approach in post-processing. As part of our in situ analysis pipeline, we employ a new method for tracking halo substructures, introducing the concept of subhalo cores. The production of multi-wavelength synthetic sky maps is facilitated by generating particle light cones in situ, also beginning at $z~\sim 10$. We provide an overview of the simulation setup and the generated data products; a first set of analysis results is presented. A subset of the data is publicly available.

\end{abstract}

\keywords{methods: N-body ---
          cosmology: large-scale structure of the universe}}


\twocolumn[\head]

\section{Introduction}

Cosmology has undergone a sea change over the last three decades. Remarkable improvements in survey observations have changed many aspects of the field from being qualitative in nature, and even speculative, to the present era of ``precision cosmology''. The current cosmological standard model, $\Lambda$CDM, successfully describes all observations using only a handful of parameters, each determined at an accuracy level of a few percent (for recent parameter constraints, see, e.g., \citealt{planck18}). Although $\Lambda$CDM is phenomenologically very successful, its two key ingredients, dark matter and dark energy, remain mysterious. Additionally, the origin of primordial fluctuations requires further investigation.

In order to probe deeper into the unknown, major ground- and space-based surveys are being constructed and proposed. Among ground-based surveys, DESI~\citep{desi} recently achieved first light, and the Vera C. Rubin Observatory is readying to carry out LSST~\citep{lsst,desc}. From space, surveys such as WFIRST~\citep{wfirst}, Euclid~\citep{euclid} and SPHEREx~\citep{spherex} promise major advances in obtaining a host of cosmological measurements. With the advent of these surveys, the quality of the observations and their statistical completeness have led to an increased and intense focus on understanding systematic errors in all aspects of the inference chain that cosmology, as an observational science, must employ. Major modeling and simulation challenges have to be overcome to enable systematic and robust exploration of the scales covered by the observations. 

\begin{figure*}[t]
\centerline{\includegraphics[width=6.2in]{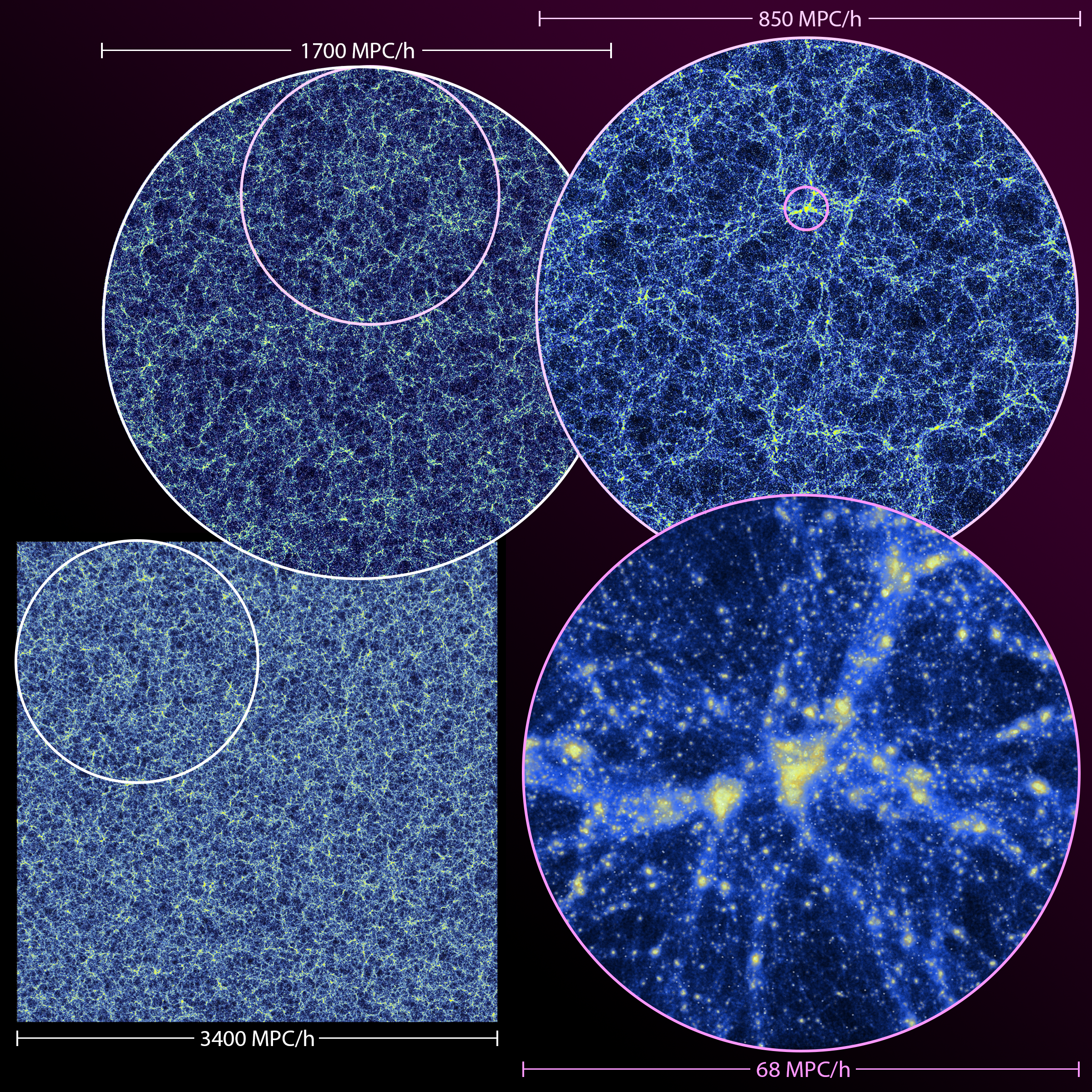}}
\caption{\label{fig:vis} Visualization of the Last Journey simulation. Shown are thin density slices for the full box (lower left corner) and zoom-ins at different levels. The panel on the lower right focuses on the largest cluster in the simulation with a mass of $\sim 6\cdot 10^{15}h^{-1}$M$_\odot$.} 
\end{figure*}

The requirements for simulations that cover the size and depth of upcoming cosmological surveys are very demanding -- only a handful of such simulations have been carried out so far. The volume required to both enable the creation of light cones out to high redshift and to properly sample rare cluster-mass objects is approximately (4-5Gpc)$^3$. In order to capture galaxies with luminosity $\sim 0.1L^{\star}$, halo masses of order $10^{11}h^{-1}$M$_\odot$ must be adequately resolved. This sets a requirement on the mass resolution of the simulation of
$m_p\sim 2\cdot 10^9h^{-1}$M$_\odot$ or better and, by itself, implies the need for simulations that evolve trillions of particles. Galaxies below the luminosity thresholds of such simulations need to be separately modeled, as described, for example, in~\cite{cosmoDC2}.

For the Euclid survey, such a simulation was carried out with the tree code {\sc PKDGRAV3} on the Piz Daint supercomputer, a GPU-enhanced system. The Euclid Flagship simulation~\citep{potter} evolved 12,600$^3$ particles in a (3780$h^{-1}$Mpc)$^3$ volume, with a mass resolution of $2.4 \cdot 10^9 h^{-1}$M$_{\odot}$. Another simulation of similar size is the Outer Rim run~\citep{heitmann19b}, carried out with the Hardware/Hybrid Accelerated Cosmology Code (HACC) on the IBM BG/Q system Mira in 2014. In that simulation 10,240$^3$ particles in a (3000$h^{-1}$Mpc)$^3$ volume were evolved, leading to a mass resolution of $1.85 \cdot 10^9$ $h^{-1}$M$_{\odot}$. The Outer Rim simulation~\citep{heitmann19b} has been used to generate synthetic skies for surveys such as LSST~\citep{cosmoDC2}, the South Pole Telescope (SPT,~\citealt{bleem19}), and eBOSS~\citep{eboss1,eboss2,eboss3,eboss4,eboss5,eboss6,eboss7}. There are many other cosmological simulations available, occupying useful niches in the trade-off space of mass resolution and cosmological volume, but there are relatively few that simultaneously have adequate mass resolution and simulation volume to be successfully employed in modeling contemporary surveys.

In this paper, we describe the `Last Journey' simulation, a new addition to the set of large-scale structure simulations that are specifically targeted at survey science. As with the Outer Rim simulation, the Last Journey run was carried out on Mira, a supercomputer at the Argonne Leadership Computing Facility (ALCF). Mira, a 10PFlop system, belongs to the family of IBM BG/Q supercomputers and went into production in 2012. The machine was retired early in 2020, and the Last Journey simulation was its final full-machine run. The simulation was performed with HACC, described in detail in~\cite{habib14}. The HACC framework has run very successfully on the BG/Q architecture; the Gordon Bell finalist paper in 2012~\citep{habib12} describes achieving a sustained performance of close to 14PFlops on Sequoia, another large BG/Q system. 

The Outer Rim simulation was one of the first extreme-scale simulations run on Mira, evolving more than a trillion particles and employing two-thirds of the machine. In this context, ``extreme-scale" refers to those simulations that occupy a major fraction of a system that is among the largest supercomputers worldwide. Experience with the Outer Rim led to many optimizations in the HACC framework. These optimizations were applied to I/O strategies, memory management, time stepping, and improved analysis tools. In the end, essentially every part of the code was touched and enhanced in some way. For the Last Journey run, we focused on a few additional improvements to the analysis suite. Given that Mira was on the floor for only a few more months with limited storage space availability, most of the analysis had to be carried out in situ, while the code was running on the system. This approach avoids having to store petabytes of data on the file system and also saves I/O time. In particular, we focused on speeding up the halo center finding algorithm, generating light-cone particle files in situ, and adding a range of halo properties to the output that were not measured for the Outer Rim simulation. We sought input from the LSST Dark Energy Science Collaboration (DESC) working groups and others to enable as many science projects related to major surveys as possible. 

Over the years, we have developed a customized approach to the analysis that accompanies HACC simulations to ensure high performance and excellent scalability of our in situ tools. Our parallel analysis tool suite, called CosmoTools, takes advantage of the data layout of the main code and reuses optimized HACC algorithms, such as the large-scale distributed FFT\footnote{A stand-alone version of HACC's distributed-memory, pencil-decomposed, parallel 3D FFT, SWFFT, is available at the following URL: https://xgitlab.cels.anl.gov/hacc/SWFFT}, wherever possible. CosmoTools is triggered at selected time steps via a separate input deck in which the analysis tools and their parameters (e.g., the friends-of-friends (FOF) halo finder linking length) are specified. We stress that it is not possible to straightforwardly integrate a number of commonly used community tools and to scale them in a performant manner up to the run sizes needed for a simulation like the Last Journey. Therefore, the continued development of CosmoTools has been a crucial step in enabling the work described in this paper. As an example of the analysis output from CosmoTools, Figure~\ref{fig:vis} shows the result of a zoom-in to the largest halo in the Last Journey with a mass of $m_{h}\sim6\cdot10^{15}h^{-1}$M$_\odot$ measured with an FOF halo finder using a linking length of $b=0.168$, shown in the lower right corner of the figure.

A major goal for the Last Journey simulation is to provide a base data set for generating detailed synthetic sky catalogs. The generation of such catalogs is a highly nontrivial task, and no single method has proven to be decisively superior over others so far. 
For the Outer Rim simulation, we employed a hybrid approach by combining empirical and semi-analytic modeling (SAM), with a much larger weight given to the empirical modeling, resulting in the cosmoDC2 extra-galactic catalog~\citep{cosmoDC2}. One reason we took this route was that Galacticus~\citep{benson}, the SAM we chose to use, is very compute-intensive and therefore difficult to carefully tune to match observational constraints. For the Last Journey simulation, we plan to use {\sc L-Galaxies}\footnote{http://galformod.mpa-garching.mpg.de/public/LGalaxies/} (see~\citealt{lgalaxies} and references therein), another publicly available SAM with somewhat less reliance on empirical models. This work will be described in a future paper. In addition to generating galaxy catalogs, we will also compute weak lensing information to provide a complete set of synthetic maps for dark energy studies with optical surveys and a range of data products geared toward ongoing and upcoming cosmic microwave background (CMB) experiments, which will also facilitate cross-correlation studies. Throughout the paper, we will highlight the specific data products generated to enable these goals.

The paper is organized as follows. In Section~\ref{sec:sim} we describe the simulation specifications, including the cosmology, simulation volume, and mass resolution. First level outputs from the simulation are then described in Section~\ref{sec:outputs}. Next, we list the data products that are obtained in post-processing in Section~\ref{sec:postproc}. We show selected results in Section~\ref{sec:results}, including measurements of power spectra, mass functions, and the halo concentration-mass relation. The results are in good general agreement with previous work, confirming the quality of the simulation. As part of this paper, we release some of the data products and describe those in Section~\ref{sec:release}. Finally, we present a conclusion and provide an outlook in Section~\ref{sec:summary}.

\section{Simulation Specification}
\label{sec:sim}

The Last Journey simulation evolved 10,752$^3$ particles ($\sim 1.24$ trillion) in a (3400$h^{-1}$Mpc)$^3$ = (5025Mpc)$^3$ volume. We have chosen a best-fit Planck 
cosmology~(\citealt{planck18}, Table 2, base-$\Lambda$CDM fit combining CMB spectra, with CMB lensing reconstruction and baryon acoustic oscillation (BAO) measurements), a $\Lambda$CDM model with massless neutrinos:
\begin{eqnarray}
\Omega_{\rm cdm}&=&0.26067, \\
\omega_b&=&0.02242 \Rightarrow \Omega_b=0.049, \\
h&=&0.6766, \\
\sigma_8&=&0.8102, \\
n_s&=&0.9665, \\
w&=&-1,
\end{eqnarray}
resulting in a total matter contribution of $\Omega_m=0.310$. This leads to a mass resolution of $m_p\sim2.7\cdot 10^9 h^{-1}$M$_\odot$. The simulation was started at $z_{\rm in}=200$ using the Zel'dovich approximation~\citep{Zel70} for setting up the initial conditions. We used the publicly available code {\sc CAMB}~\citep{camb} to generate the input transfer function. A detailed description of the initial condition generation is given in~\cite{Heitmann:2010}. In the same paper, the use of the Zel'dovich approximation versus higher-order schemes is discussed in detail and carefully tested. It was found that that if the simulation is started at a high enough redshift and in a large volume such that the initial move of the particles is a fraction of the initial particle inter-spacing, the Zel'dovich approximation provides highly accurate initial conditions, basically indistinguishable from higher-order schemes. In addition, a high redshift start ensures that structures have enough time to form correctly at the scales of interest. We followed the findings in~\cite{Heitmann:2010} to set up very accurate initial conditions for the Last Journey simulation. The simulation was carried out with HACC running in its tree-PM mode, with a force resolution of $3.16h^{-1}$kpc. The high mass and force resolution in combination with the large volume will enable the creation of detailed synthetic sky catalogs across wavebands, including weak lensing maps, cluster weak lensing maps, optical cluster modeling, and galaxy clustering, to name just a few. A wide range of different modeling approaches can be employed for the generation of such catalogs due to the excellent resolution provided by the Last Journey simulation.

\section{Outputs}
\label{sec:outputs}

Outputs from the HACC simulation consist of raw particle information (full and subsampled) as well as derived quantities. In order to reduce the amount of data generated, we employ two strategies: first, we compress the data in a lossless fashion using the Blosc library\footnote{https://blosc.org/} and second, we carry out as much of the analysis as possible in situ to avoid storing raw data files. The compression of particle files that contain mostly positions and velocities leads to a reduction of data by a factor of 1.4. For outputs that are dominated by integers (e.g. files that contain halo and particle tags), the compression factor can be as high as 7$\times$. The compression rates are at the expected level, as detailed in the following. The particle data comprise 36-bytes per particle, 12 bytes for the velocities (which do not compress very well), 12 bytes for positions (for which the sign and
exponent will compress well and the lower mantissa bits will not, leading to a target of 72\% compressed size),
and the remainder is essentially metadata (particle IDs, mass, mask) that almost completely compress. Thus, ideally, we would obtain an overall 57\% compression rate for the particle files via $(12+12\times 0.72)/36$. The
mostly regular information does not compress perfectly (because there
is some information content), and the compressor is not perfect, so $1.4\times$
is in the expected range. The total amount of analysis data generated was $\sim$720TB. We also stored five full snapshots ($\sim$~150TB) of the full raw particle data for possible post-analysis studies. 

As in previous large-scale simulations carried out with HACC, we store outputs at 101 time snapshots between $z=10$ and $z=0$, evenly spaced in
$\log_{10}(a)$ for most of our data products. This leads to the following output values in redshift:
\begin{eqnarray}
z&=&\left\{10.04, 9.81, 9.56, 9.36, 9.15, 8.76, 8.57, 8.39, 8.05,
   \right.\nonumber\\ 
&&7.89, 7.74, 7.45, 7.31, 7.04, 6.91, 6.67, 6.56, 6.34, 6.13, 
\nonumber\\ 
&&6.03, 5.84, 5.66, 5.48, 5.32, 5.24, 5.09, 4.95, 4.74, 4.61, 
\nonumber\\ 
&&4.49, 4.37, 4.26, 4.10, 4.00, 3.86, 3.76, 3.63.  3.55, 3.43, \nonumber\\ 
&&3.31, 3.21, 3.10,
3.04, 2.94, 2.85, 2.74, 2.65, 2.58, 2.48, 
\nonumber\\ 
&&2.41, 2.32, 2.25, 2.17, 2.09, 2.02, 1.95, 1.88, 1.80, 1.74, 
\nonumber\\ 
&&1.68, 1.61, 1.54, 1.49, 1.43, 1.38, 1.32, 1.26, 1.21, 1.15, 
\nonumber\\ 
&&1.11, 1.06, 1.01, 0.96, 0.91, 0.86, 0.82, 0.78, 0.74, 0.69, 
\nonumber\\ 
&&0.66, 0.62, 0.58, 0.54, 0.50, 0.47, 0.43, 0.40, 0.36, 0.33, 
\nonumber\\ 
&&0.30, 0.27, 0.24, 0.21, 0.18, 0.15, 0.13, 0.10, 0.07, 0.05, 
\nonumber\\ 
&&\left.
0.02, 0.00\right\}.
\label{redshifts}
\end{eqnarray}

\subsection{Snapshots}

We save downsampled particle outputs at the 101 redshifts specified above. These snapshots can be used to generate light cones with different observer positions, estimates of correlation functions at specific redshifts, and coarse-grained density maps at specific redshifts. We randomly save 1\% of the total number of particles in each snapshot where the choice of selected particles is kept the same after the initial snapshot. Overall, the storage requirement for these files is $\sim$30TB. In addition to the downsampled particles, we also saved a small number of full outputs at the following redshifts: $z={2.02, 1.01, 0.58, 0.15, 0.0}$. The storage requirement for these files is $\sim$150TB. 

\subsection{Particle Light Cones}

Below redshift $z=10$, particle light cones are generated and output during each simulation time step. We store a full sphere relative to an observer placed in the corner of the simulation box at $(0,0,0)$ for the generation of full-sky maps and realizations of smaller area outputs as required. For the redshift range, $3\le z\le 10$, the outputs are saved as light-cone particles downsampled to 1\% of the full particle count. For redshifts $z<3$, we save full particle light cones. In addition to positions and velocities, we store the potential from the particle mesh (PM) solver interpolated to the particle position on the light cone. This information will enable the reconstruction of a (coarse-grained) tidal field to, e.g., carry out future intrinsic alignment (IA) studies. The full light-cone outputs out to $z\sim 3$ result in $\sim$286TB of data, whereas the downsampled particle outputs to $z\sim10$ result in 7TB of data.

At high redshifts, the volume of the simulation is not sufficient to cover the full sphere. At redshift $z=10$, we therefore replicate the simulation volume 64 times (without rotations) to cover a comoving distance of 6519.52$h^{-1}$Mpc, equivalent to the radius of the light-cone sphere. The large memory overhead of the replication requires downsampling of the particles to enable the in situ evaluation of the light cone keeping within the computer's memory. 
The downsampling also reduces the storage footprint required for the light cone. For the science cases of interest, e.g. CMB lensing, the downsampled light cones at high redshift are sufficient to generate maps at the required accuracy. At lower redshifts, the number of replicants reduces accordingly with the shrinking light-cone sphere, allowing us to save the full particle light cone for $z<3$.

This method of in situ particle light-cone generation has several benefits over post-processing from particle outputs. It removes the need for particle outputs at every analysis step, which are expensive both in terms of storage and I/O. Additionally, it is well known that extrapolation or interpolation between analysis steps causes small-scale inaccuracies in light cones when orbital time scales are comparable to or smaller than the inter-step spacing \citep[see, e.g., ][]{cosmoDC2, merson13, kw07}. While neglecting the evolution between analysis steps can reduce these small-scale errors, it introduces discretization inaccuracies into the clustering. The in situ computation uses extrapolation of the position and velocity data within the simulation timestep integrator, effectively setting the time-resolution of the light-cone computation to be similar to that of the simulation itself. In this way, the light-cone inaccuracies and post-processing requirements are significantly reduced compared to the post-processing computation. 

\subsection{Halos} 
\label{sec:halos}
Next, we describe the halo files that are generated as part of the Last Journey simulation. They provide the foundation for the creation of synthetic sky maps.  Halo information enables detailed studies of structure formation processes and investigations of the galaxy-halo connection. We have run an FOF halo-finder and determined halo spherical overdensity (SO) properties at the 101 snapshots listed in Equation~(\ref{redshifts}). The FOF finder is based on the standard definition first introduced for cosmological applications in~\cite{davis85} and follows a fast tree-based algorithm~\citep{fof-alg}. We use the centers of the FOF halos (given by their local potential minima) to grow SO halos. This approach allows for efficient identification of SO halos and generally avoids overlaps between halos. However, one concern might be that due to the irregular shapes of the FOF halos and the linking of halos that can occur in the FOF approach, small halos might be missed. \cite{knebe11} carried out an extensive halo finder comparison project, including the halo finder that is used here (the ``LANL'' halo finder in \citealt{knebe11}). They found excellent agreement between the LANL and other SO finders if the SO halos are well sampled with at least 500 particles,\footnote{In \cite{knebe11}, a linking length of $b=0.2$ was used. The overlinking problem is less severe with a smaller linking length. In this paper, we use $b=0.168$. Together with our minimum mass of 500 particles for the SO halos, this choice ensures that we capture the SO halos reliably.} the same mass cut we employ here for the SO halos. The in situ halo information results in close to 200TB of data. We have stored the following files and information.

\begin{itemize}
\item 
Halo properties: We measure halo properties for both FOF and SO halos. The mass cut on FOF halos is 20 particles per halo, leading to a minimum halo mass of $m_{\rm FOF}\sim5.4\cdot 10^{10}h^{-1}$M$_\odot$. We use a linking length of $b=0.168$. This value is commonly used to create halo occupation distribution (HOD)-based mock catalogs for, e.g., redshift space distortion studies \citep{reid11}, measurements of the clustering of massive galaxies in the BOSS survey~\citep{white11} and several recent eBOSS studies~\citep{eboss1,eboss2,eboss3,eboss4,eboss5,eboss6,eboss7}. We find SO halos for each FOF halo that has at least 500 particles. In addition, for five snapshots between $z=0$ and $z=2.02$, we measure FOF halo properties for a linking length of $b=0.2$ to enable mass function comparisons with results available in the literature (see Section~\ref{sec:HMF} for details).
\begin{itemize}
\item
For the FOF halos, we save the number of particles in a halo, the halo tag, the halo mass in units of $h^{-1}$M$_\odot$, the kinetic energy, the halo centers as measured by the local potential minimum and the center of mass, the halo angular momentum, the halo circular velocity, the halo velocity, the halo velocity dispersion, and the eigenvectors of the halo inertia tensor.
\end{itemize}
\begin{itemize}
\item
For the SO halos, we save M$_{200c}$ measured as the number of particles in units of $h^{-1}$M$_\odot$, the halo radius ($r_{200c}$ in units of $h^{-1}$Mpc), the kinetic energy, the velocity dispersion, the circular velocity, the angular momentum, the halo velocity, the inertia tensor eigenvectors, and the concentration measured in three different ways (profile fitting, accumulated mass and peak measurement, see~\cite{child} for details).
\end{itemize}
\item 
Halo particle tags: The halo particle tag files contain tags for all particles that reside in halos and their halo tag. The main use case for this output is the construction of merger trees in post-processing.
\item
Halo particles: We store all particles in halos that have at least 10,000 particles, translating into a halo mass of $m_h=2.7\cdot 10^{13}h^{-1}$M$_\odot$. We also store randomly selected particles from all halos downsampled at a rate  of 1\%, with at least five particles per halo. 
\item
SO mass profiles: For all SO halos, we store their mass profiles binned in 20 radial bins.  
\end{itemize}

\subsection{Halo Cores}
\label{sec:halocores}

In addition to identifying halos and determining their properties at each analysis time step, we also save halo ``core'' information over time. Halo cores, as described further below, enable tracking of halo substructure information.
For each FOF halo with at least 80 particles, we store the halo core, defined here as the 50 particles closest to the potential center of the halo. These particles are saved in core particle files for each of the 101 analysis timesteps. The raw files add up to $\sim$32TB of storage. Additionally, once a particle has been identified as a core particle, it is tracked and recorded until the last analysis timestep, independently of whether it is in a halo or not in the next analysis step. These accumulated core files add up to almost another 150TB. The core and accumulated core files allow us to track halo substructure by following halo core particles that have fallen into more massive host halos, serving as an alternative to subhalo finding and tracking. In a post-processing step, we generate a core catalog as described in Section~\ref{sec:CCG}. The core catalog contains information about the cores themselves, such as sizes, positions, and velocities, as well as their temporal history and later role inside a halo (central versus satellite) and the halo properties at infall (when merging into a larger halo). Once the core catalog is generated, the core and accumulated core files do not need to be stored.   

\section{Post-processing Analysis}
\label{sec:postproc}

From the information saved during the in situ analysis steps described in Section~\ref{sec:outputs}, several additional data products are generated in post-processing. These data products are used to build synthetic sky catalogs in different wavebands and to track the evolution of structure over time. The approach we are taking is customized for the HACC analysis infrastructure to ensure the required scalability and speed needed when analyzing very large simulation outputs. In the following, we provide a brief description of each of the data products and their potential usage in follow-up science investigations.   

\subsection{Merger Trees}

Merger trees describe the hierarchical formation of halos and subsequent mass accretion by computing the overlap in halo particles from adjacent catalog snapshot files. They provide the means to gain a more fundamental understanding of the hierarchical nature of structure formation, as well as a framework within which to model galaxy formation in the cosmological context. In particular, merger trees are essential for investigating the galaxy-halo connection if one desires to go beyond simpler approaches based only on individual halo properties (for a recent review of the galaxy-halo connection see \cite{wechslertinker18} and references therein). The development of synthetic sky catalogs using methods that rely on detailed information of halo and substructure evolution, such as SAMs, is also based on merger trees. Merger tree construction is a nontrivial process, and different methods have been developed in the literature (for a comparison study and description of popular merger tree codes, see \citealt{srisawat13}).

Challenges in merger tree building include mass-resolution effects and halos that ``split", i.e. halos that may have multiple descendants in subsequent snapshots. Finite mass resolution (and the connected minimum halo mass) can have a stochastic thresholding effect leading to halos that disappear from one snapshot to the next. This problem has long been recognized, and different mitigation schemes have been discussed in the literature for both halo and subhalo merger trees (see, e.g., \citealt{fakhouri08}, \citealt{behroozi13}, \citealt{sublink}, and \citealt{han18} and references therein). Instead of inserting `phantom' halos to account for disappearances, we mitigate this problem by using a soft threshold. Once a halo has been found because the particle count crossed the desired threshold (e.g., 50 particles), we can allow that halo to continue to exist in subsequent analysis steps even if its mass dips below the original preset value to some new minimum value (e.g., 20 particles). 

Halo splitting commonly occurs in simulations, as neighboring FOF halos can occasionally be overlinked and identified as a single object, even when they have not dynamically merged. In later time steps these halos may be rediscovered as separate objects. Moreover, some of these halos are involved in fly-bys and never merge after splitting. These complications can make halo temporal connectivity ill-defined. To address splitting, we utilize a halo `fragmentation' approach described below, intended to track all identified mass objects in the simulation by artificially breaking up halos that are later identified to have split. 

Our merger tree implementation is focused on good scalability, efficient memory usage, and high accuracy. A detailed description of our approach, with an emphasis on algorithmic details, is given in~\cite{rangel17}; we provide a brief summary here. We stress that we only describe the construction of halo merger trees in this section and not subhalo merger trees, commonly used for the analysis of high-resolution simulations. To track substructures (structures that have fallen into a halo), we utilize our newly developed core-tracking approach, introduced in~\cite{rangel17} and applied to the modeling of galaxy distributions in~Korytov et al. (2020, in preparation). We summarize the core methodology later in Section~\ref{sec:CCG}.

To facilitate building merger trees, we store the tags of all particles that reside in halos and the corresponding halo tag at each of the 101 analysis steps. In addition, we store all halo properties in the halo catalog, queryable using the same halo tags. This information allows us to track the evolution of all halos via their particle content. The construction of the merger trees starts with the last snapshot at $z=0$ and then works backward in time until the first analysis snapshot is reached at $z\approx 10$. We identify overlapping halo particles between adjacent snapshots to determine the progenitors of each descendant halo. If a halo has an overlapping particle set above the minimum halo mass with multiple descendant halos (indicating the halo had split), we break up the halo into corresponding fragments. The original halo mass is assigned to the fragments proportionally by taking the ratio of each fragment intersection and their union (conserving total mass). As a result, the descendant of every halo is the halo with the greatest amount of particle overlap. 

The fragment halos serve as markers for mass objects with individual properties not captured in the catalog, as their existence was hidden due to numerical effects such as overlinking. We note the advantage of temporally constructing the tree in reverse time. All of the objects at $z=0$ serve as the base halos (final descendants) of every tree. Walking backward we can connect halos and construct artificial fragments as needed in a connectivity tree that consistently tracks all mass that eventually ends up in the final objects of interest at $z=0$. Flyby halos, by construction, will not appear in the final tree, as they never ended up in the final mass at $z=0$. Constructing trees forward in time would require continuously breaking up trees during every split encountered and can significantly complicate merger tree construction (often requiring multiple passes of tree pruning to correct). In summary, our approach produces clean merger trees, where all halos at any given redshift have one descendant (no splits), and for each halo, we have amassed its entire connectivity history. 

\subsection{Core Catalog Generation}
\label{sec:CCG}

\begin{table*}[t]
\caption{Properties in the core catalog file}
\begin{tabular}{ll}
\hline\hline
Property & Description\\
\hline\hline
$x$, $y$, $z$ &  Position of central core particle  \\
$v_x$, $v_y$, $v_z$    &   Velocity of central core particle   \\
Radius &  Core radius \\
Velocity dispersion &  Velocity dispersion of core particles \\
 \hline
 Core tag &  Unique label of core (assigned at core creation and kept unmodified throughout catalog)  \\ 
 Tree node index &  Unique identifier of merger tree object containing core at current timestep   \\
 Central/satellite &  1 if a central core, 0 if a satellite core. Central cores are assigned properties of the host merger tree object \\
 Host core tag & Core tag associated with central object at time of infall (same as core tag for centrals) \\
\hline
 Infall step &  Simulation timestep immediately preceding infall (current step if central)\\
 Infall halo properties  & All host halo quantities recorded at infall; for centrals, it is the current step halo properties \\
 \hline\hline
 \end{tabular}
\label{tab:properties}
\end{table*}

The core catalog is generated from three input data sets: the core particle files, the accumulated core files, and the merger trees. Incorporating the merger tree information allows us to track not only halos but also cores over time in the core catalog. We also include information from the halo property files for each core. Table~\ref{tab:properties} summarizes the information stored in the core catalog. In many ways, the core catalog is the main analysis output from the simulation, in that it encapsulates the history and the properties of each halo in the simulation. In addition to the merger tree data, it provides information about the halos after they have fallen into other halos and therefore can be substituted as a proxy for a subhalo catalog. 

Subhalo catalogs are used for a wide range of investigations, including the study of substructure in cluster-sized halos in gravity-only simulations (e.g., \citealt{gao12}) and hydrodynamic simulations (e.g., \citealt{nagai05}) or as proxies for galaxy positions (see, e.g., the investigations in \citealt{wetzel09}). In combination with merger tree information, they are the base ingredient for SAMs and subhalo abundance modeling approaches. \cite{onions12} carried out an extensive subhalo finder comparison project and found good agreement between different approaches. The subhalo finding approach comes with several associated challenges. First, for a simulation of the size of the Last Journey run, algorithms that go beyond simple spatial analysis are expensive. In particular, given that the subhalo finder has to be run on the order of 100 snapshots, the computational burden is considerable. Second, the demands on mass resolution are very severe for enabling accurate subhalo tracking. Third, the distribution of subhalos close to the center of the parent halo does not allow for correct placement of galaxies. This is due to the disruption of subhalos when they are close to the halo center. In order to overcome these challenges, we generate core catalogs that enable the tracking of the evolution of particles that at some point belonged to the inner core of a halo. This approach is a natural extension of the ideas in~\cite{kaiser84} and \cite{white87} and is related to~\cite{hong16} where the most-bound particle in a halo was tracked and used to build merger trees. The point of using multi-particle cores (as enabled by high mass resolution simulations) is to provide a potentially robust way to study the evolution of halo substructure over time.

A detailed description of the core tree assembly is given in Korytov et al. 2020 (in preparation). (The core catalog contains all of the core trees.) Here we provide a brief summary. As explained in Section~\ref{sec:halocores}, we identify in each analysis step the 50 particles that are closest to halo centers for halos with more than 80 particles and continue to track their evolution over time. For each core extracted from these files, we measure its properties, including position and velocity (taken from the centermost particle of the core), radius (defined as the root mean square of the positions), and velocity dispersion.  

Cores are marked with multiple tracking identifiers. First, we report a ``core tag'', a unique ID that does not change over time once it is assigned (i.e. when a core was first identified). The core tag is based on the merger tree node index of the halo associated with the newly birthed core. 

Next, we note whether a core is a central core or if it is a satellite; the latter terminology indicates that the core host halo merged with another larger halo, after which the core of interest is considered to be a satellite in the new halo. While constructing the core catalog, every time halos are recorded as merged in the merger tree, the absorbed halos will have their central cores converted into satellites, with the descendant halo inheriting its central core from the most massive progenitor. Moreover the satellites that existed prior to the merger in all of the progenitors will be carried into the merged descendent as well; thus, a given halo at any snapshot only has one central core, with the rest of its substructure encapsulated by all of the satellites it has accrued over all of its previous merger history. 

Note that, unlike the merger tree, which was constructed backward in time, the core trees are created forward in time, and track core history via the discussed merging criteria. While building the core trees, the central core properties are extracted from the core particle files, as they are measured for every halo on every analysis snapshot. Satellite halo properties are determined separately. As all satellites were originally centrals of a halo that merged, during catalog construction, we temporarily store the central core particles of all merged halos at infall. Using this stored particle set, satellite properties at any following snapshot post-infall are then extracted from the accumulated core particle files, which we recall have been tracking up-to-date properties of any particle ever identified as being in a core.  

The second core tag we record in the catalog is referred to as the ``tree node index", which indicates the hosting halo of the core in the given snapshot. If the core is a central, the core tag and the tree node index are the same. If the core has merged into another halo and, therefore, is a satellite, the tree node index will indicate the new parent halo. The final identifier is the ``host core tag". This tag is important when a core falls into another halo -- the host core tag then reports the core tag associated with the central object of its new host halo at the time of infall (a useful quantity to determine the hereditary order of a core as described in Section~\ref{sec:infall}). For a core that remains a central throughout its history, either because it lives an undisturbed existence or because it is the core of a dominant halo that absorbs other halos but is never absorbed itself, all the tags mentioned so far will be identical. 

We note that we include all halo properties in the core catalog directly to avoid having to perform matching procedures across files later. The overhead in storage is negligible. It is very important to carefully record halo properties at the time of infall for satellites (for centrals, these are simply their host halo properties). These properties can then be used to model the further evolution of the core inside its new host halo. For example, a simple mass-loss model can be applied to enable the use of cores in place of subhalos. We store all the halo properties that are available and described in Section~\ref{sec:halos}. 

To summarize, all objects in the core catalog are grouped together with the tree node index (referencing merger tree objects) at the current step, and can be traced in time using the core tag. The properties of the merger tree object itself are assigned to the central core, which in turn is marked by the central flag and has an infall step set to the current step. With some additional modeling, the information stored in the core catalogs enables the creation of substructure merger trees and, as a result, can be used as input for SAMs. We will discuss this idea further in a forthcoming paper.   

\subsection{Halo Light Cones}

Dark matter-dominated halos form the base for synthetic galaxy catalogs, in which galaxies are typically placed according to either a profile or measured substructure. Halo light cones allow for the generation of galaxy catalogs in the frame of an observer. Alongside lensing information from particle light cones these can be used to create synthetic observations of galaxy surveys such as that of \citet{cosmoDC2}. These require halo catalogs measured at regular steps throughout the simulation, as well as merger tree information on the temporal evolution of the halos. 

We generate halo light cones in post-processing from the halo catalog files and the merger trees. We select merger tree halo objects from a given snapshot and interpolate their positions backward onto the light cone using their progenitor position at the previous snapshot. For objects with multiple progenitors, we choose the most massive progenitor of the halo for the progenitor position, and when no progenitors are found we instead extrapolate the position backward onto the light cone using the velocity at the later snapshot. The case of fragment halos is treated separately -- for each set of halo fragments, we retain the most massive fragment and discard all other objects associated with the FOF halo, again using the merger tree information to interpolate or extrapolate as above. Lastly, we match against the halo catalog at the later snapshot to obtain the properties for the FOF halo associated with this fragment object, including the corresponding SO quantities as well. 

Following the procedure for the particle light cones, the halo light cones are computed for a full sphere relative to an observer placed in the corner of the simulation box at $(0,0,0)$, thus enabling full-sky maps or several realizations of smaller area outputs. These are saved for the redshift range $z<10$ with all SO and FOF halo properties as listed above. These halo light cones are used to generate survey-like catalogs from the simulations, as well as to access dark matter halo information in light cone coordinates. 


\section{Selected Results}
\label{sec:results}

In the following, we show selected results obtained from the Last Journey simulation. We emphasize that more results will be published in forthcoming, focused papers. In order to demonstrate the high quality of the simulation, we display a set of standard measurements compared to fitting functions and emulators available in the literature. We provide results based on particle snapshot and light-cone outputs, measurements from the halo catalogs, and finally, results based on the core catalog. 

\subsection{Matter Power Spectra}
\label{sec:powspec}

\begin{figure}[th]
\centerline{
 \includegraphics[width=3.5in]{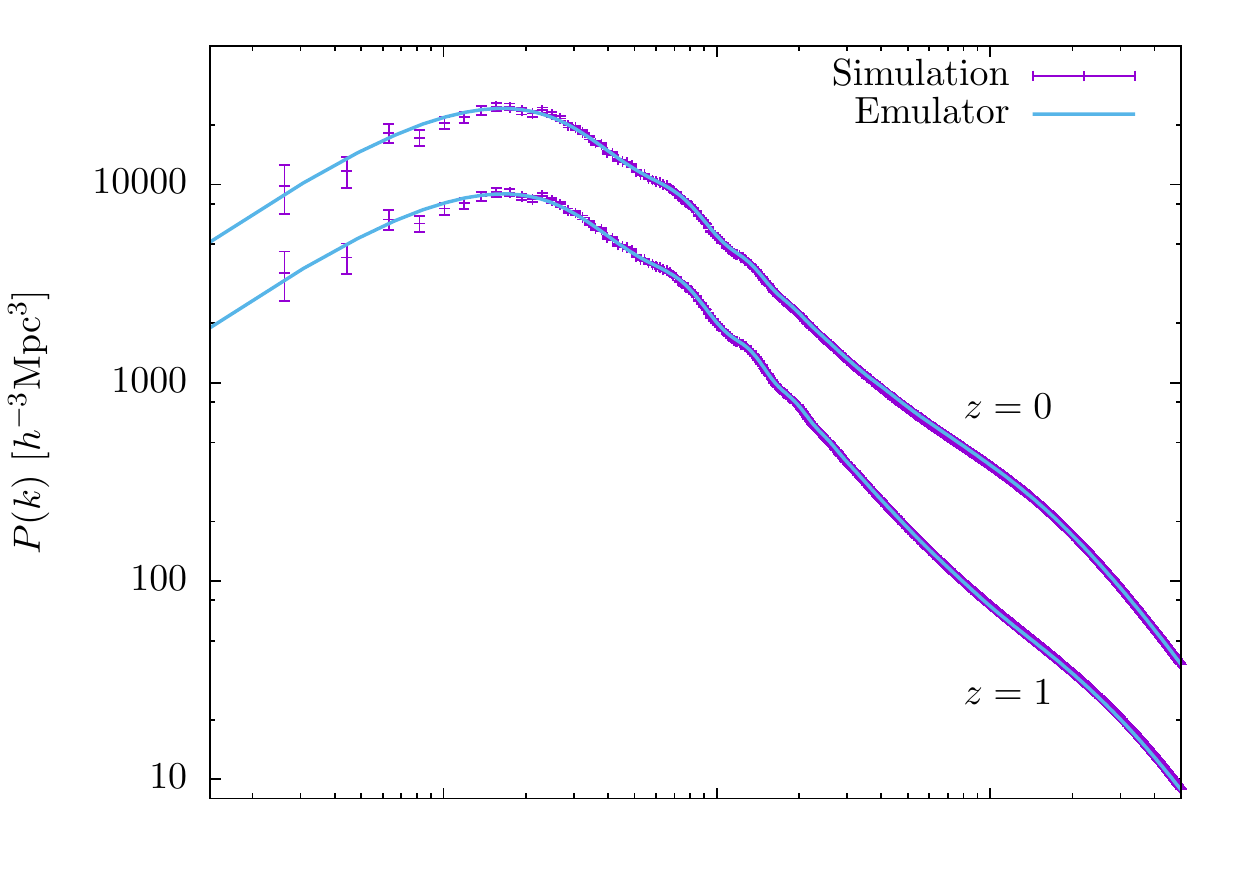}}
\vspace{-0.54cm}
\hspace{0.15cm}\centerline{\includegraphics[width=3.43in]{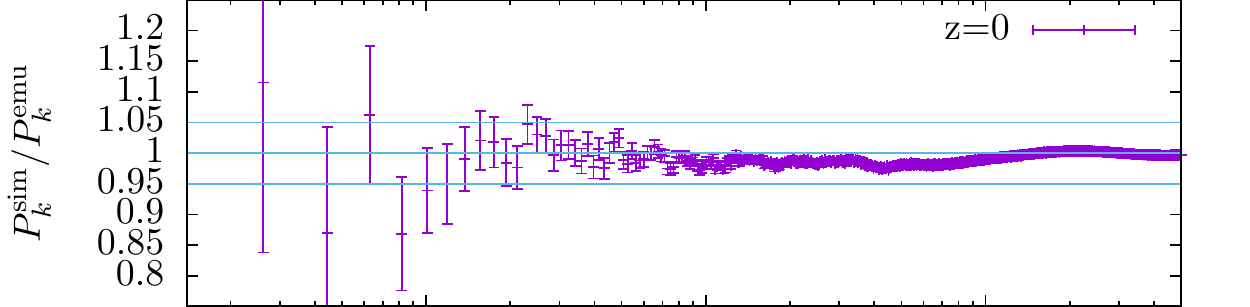}}
\vspace{-0.9cm}
\hspace{0.15cm}\centerline{\includegraphics[width=3.43in]{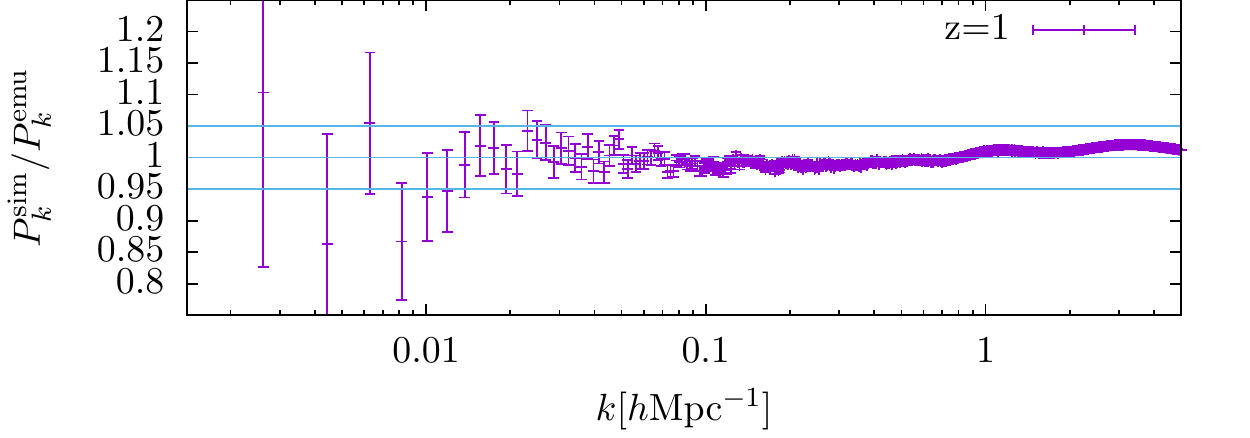}}
\vspace{0.5cm}
\caption{\label{fig:pow} Upper panel: Power spectrum results at redshift $z=0$ and
  $z=1$. For comparison, we show predictions from the Cosmic Emulator
  by~\cite{emu_ext}. Lower panels: Ratio of the simulation and the emulator at $z=0$ and
  $z=1$. The light blue bands show a 5\% range. The results are within the accuracy bounds reported in~\cite{emu_ext}.} 
\end{figure}

The nonlinear matter power spectrum is an important statistic for conveying cosmological information. From a simulation perspective, the matter power spectrum also provides a very good estimator for the accuracy of the simulation down to a certain length scale. Extensive work has been carried out in the past to compare power spectra from different codes \citep{Heitmann:2005,Heitmann:2007hr,schneider16} and to assess the accuracy at which the matter power spectrum can be obtained given certain simulation settings, such as volume, particle loading, and force resolution~\citep{Heitmann:2010}. The HACC framework automatically generates a power spectrum measurement  every time it stores a checkpoint/restart file. Overall, we saved 55 power spectra from the full simulation. Figure~\ref{fig:pow} shows the measurement of the power spectrum from the Last Journey simulation at two redshifts, $z=0$ and $z=1$. In addition, we show predictions at those redshifts from the Cosmic Emulator \citep{emu_ext}. The agreement is excellent, well below 5\%, the reported accuracy for the emulator. The comparison with the emulator is also an indirect comparison of {\sc Gadget-2}~\citep{gadget2} and HACC, given that the emulator used in this comparison is based on a large suite of {\sc Gadget-2} simulations.
The error bars shown are those that would be expected for a Gaussian Random Field with the measured power spectrum,
with the standard deviation of power in the $i$th wavenumber bin, $\sigma_{P(k_i)}$, given by
\begin{equation}
    \sigma_{P(k_i)} = \frac{P(k_i)}{\sqrt{N_{\rm modes}(k_i)}},
\end{equation}
where $P(k_i)$ is the estimated power in the bin and $N_{\rm modes}(k_i)$ is the number of independent modes in the bin. These error bar estimates were checked against the variance in ensembles of discrete Gaussian Random Field realizations.

\subsection{Matter Correlation Function}

\begin{figure}[b]
\centerline{
 \includegraphics[width=3.5in]{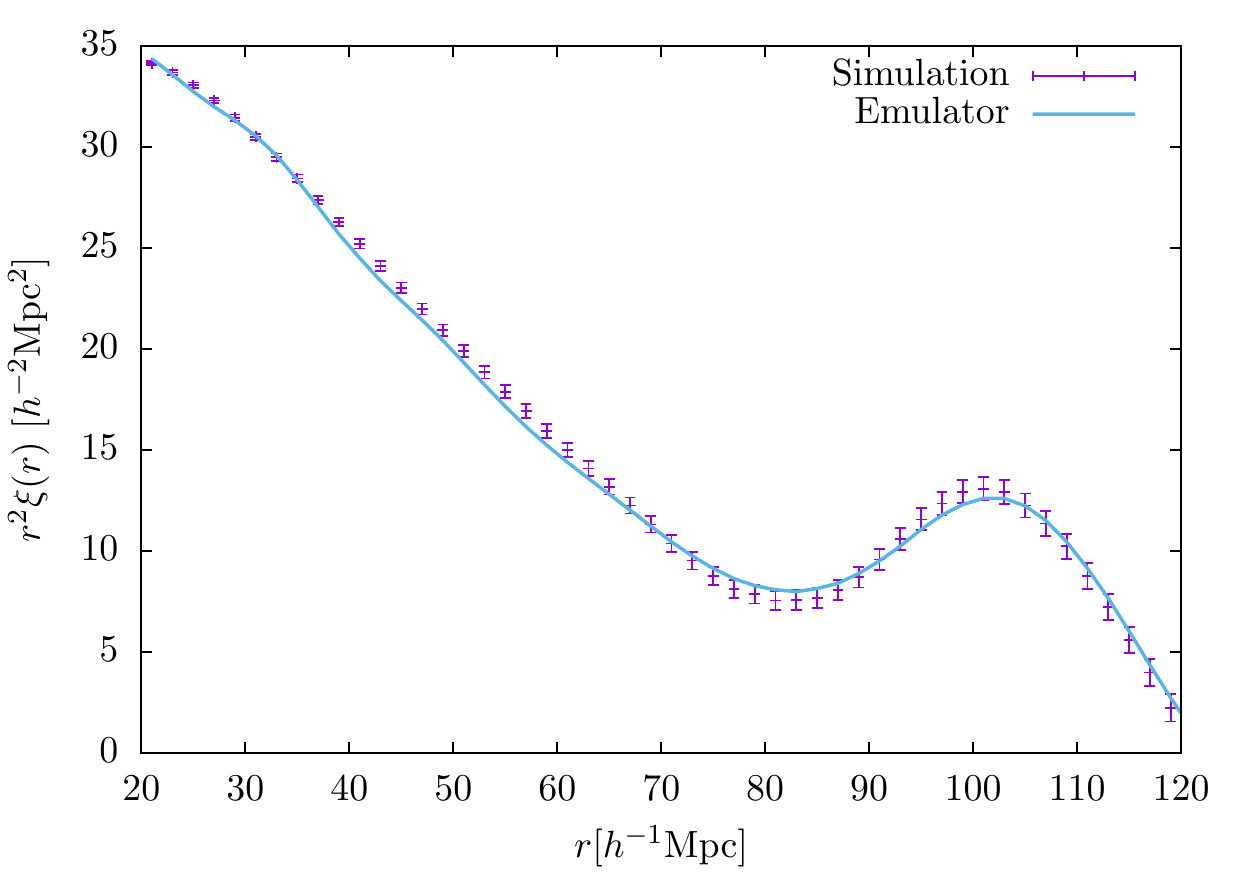}}
\caption{\label{fig:xi} Real space correlation function of downsampled particles at redshift $z = 0$ as computed from Eq.~\ref{eq:minus} and Eq.~\ref{eq:rr}. For comparison we show a prediction based on the Cosmic Emulator by \cite{emu_ext}. We leave this as a qualitative comparison as the prediction was not directly derived from measuring correlation functions  but via a Fourier transform of the power spectrum emulator, which makes the accuracy of the prediction difficult to quantify.}
\end{figure}

We also computed the two-point correlation function, the real-space analog of the power spectrum. This was performed by counting pairs on a 1\% downsampled population of particles (approximately 12 billion), counting the total number of particle pairs $DD_i$ that fall within the $i$th bin in separation distance, with the midpoint of the bin denoted by $r_i$. The correlation function is then computed using the Minus estimator
\begin{equation}
\xi(r_i) = \frac{DD_i}{RR_i} - 1,
\label{eq:minus}
\end{equation}
where $RR_i$ is the equivalent number of particle pairs that would fall within the distance bin given a random distribution (see Eq.~3.12 in \citealt{martinezsaar} or Eq.~5 in \citealt{kerscher2000}). As in our case the geometry is a periodic box, we compute this term analytically using 
\begin{equation}
RR_i = NV_i\frac{N}{L^3},
\label{eq:rr}
\end{equation}
where $L$ is the simulation box length, $N$ is the number of particles used to count pairs for the $DD$ values, and the volume of the spherical shell corresponding to the bin (of width $\Delta_i$) is $V_i = \frac{4}{3} \pi [(r_i + \frac{1}{2}\Delta_i)^3 - (r_i - \frac{1}{2}\Delta_i)^3]$.
An important result demonstrated by this calculation is the clear BAO signal at \textasciitilde{}$100~h^{-1} {\rm Mpc}$, as shown in Figure~\ref{fig:xi} at redshift $z=0$, which plots the product of the measured correlation function and the square of the bin midpoints. The location and shape of the BAO bump is consistent with the overlaid model prediction using the power spectrum of \cite{emu_ext}, converted to real-space using the {\rm mcfit}\footnote{https://github.com/eelregit/mcfit/} Python package of \cite{mcfit}.

The error bars presented here are similar in scope to the power spectrum, but error bars in correlation function space are highly correlated, and estimation is somewhat more involved. Following \citet{sanchez2008}, we start with the variance expected from the Gaussian component of density fluctuations and a shot-noise component from finite tracer density \citep{fkp94}
\begin{equation}
    \sigma_P^2(k) = \frac{2}{V} \left(P(k) + \frac{1}{\bar{n}}\right)^2,
    \label{eq:vark}
\end{equation}
where $V$ is the simulation volume and $\bar{n}$ is the mean tracer density. The covariance of the two-point correlation function is then given by \cite{cohn2006,smith2008}:
\begin{equation}
    C_\xi(r,r') = \int \frac{k^2{\rm d}k}{2\pi^2} j_0(kr) j_0(kr') \sigma_P^2(k),
    \label{eq:2ptcov}
\end{equation}
where $j_0$ is the zeroth-order spherical Bessel function, and the standard deviation of the two-point correlation function is then $\sigma_\xi(r) = \sqrt{C_\xi(r,r)}$. We used the {\rm 2D-FFTLog}\footnote{https://github.com/xfangcosmo/2DFFTLog/} Python package from \citet{2dfftlog} to compute the integrals in Eq.~\ref{eq:2ptcov}, and we verified the error bar estimates against variance of the two-point correlation functions measured from several ensembles of discrete Gaussian Random Fields. A finite bin width will dilute the covariance, but for the binning presented here (bin width of $2~h^{-1} {\rm Mpc}$ from $20~h^{-1} {\rm Mpc}$ to $130~h^{-1} {\rm Mpc}$) we found that the finite bin width reduces our error bar estimates by less than $3\%$, which is expected to be subdominant to other omitted effects, so we use Eq.~\ref{eq:2ptcov} directly in the thin shell approximation. We also found that the shot-noise component in Eq.~\ref{eq:vark} comprises less than $0.1\%$ of the errors for our density of tracer particles, and so we conclude that our downsampled particle density is sufficient to fully resolve the measurement in the simulation box.

\subsection{Angular Correlation Function}

\begin{figure}[h]
\centerline{
\hspace{-0.1cm}\includegraphics[width=3.5in]{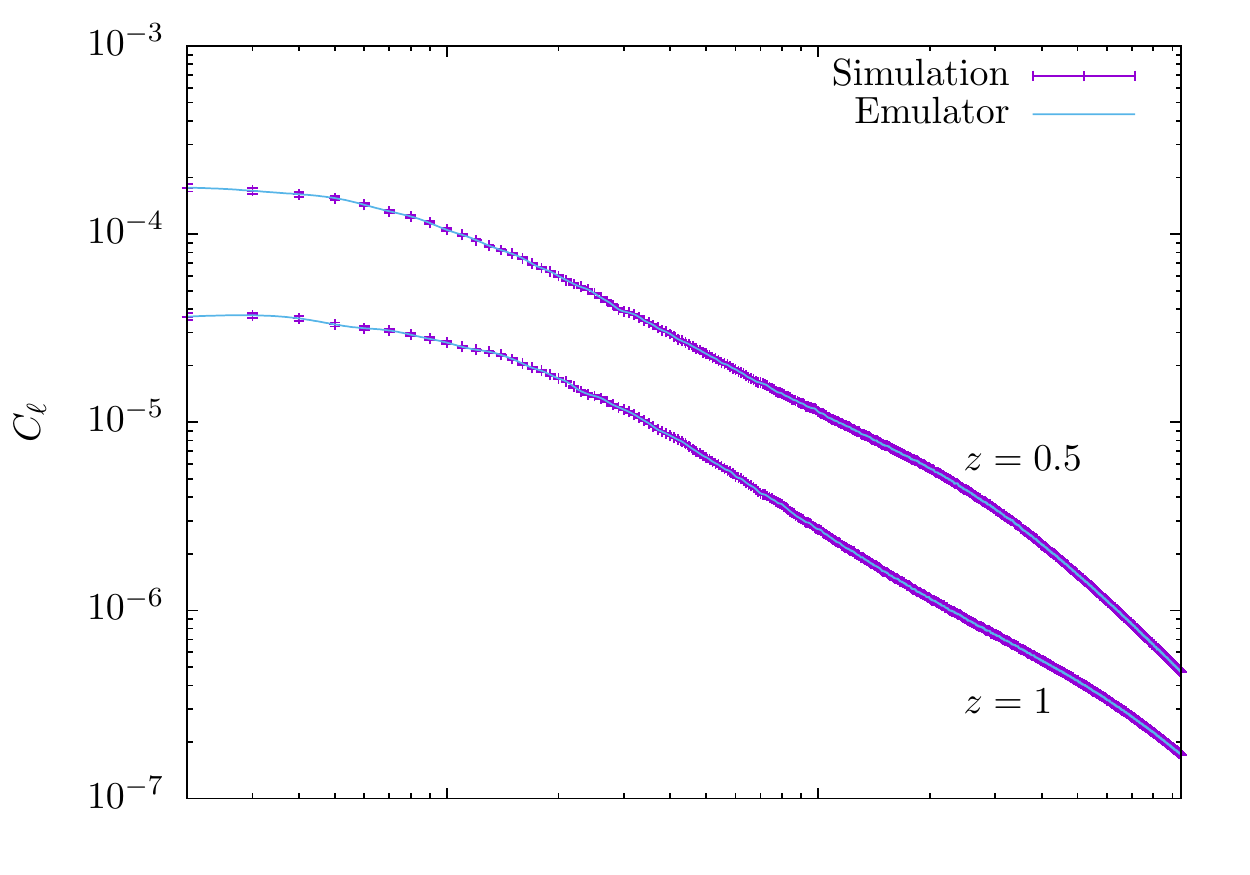}}
\vspace{-0.54cm}
\centerline{\includegraphics[width=3.5in]{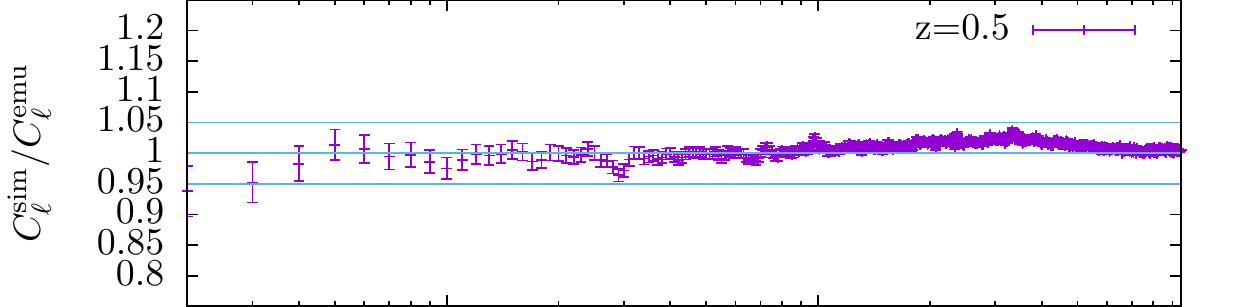}}
\centerline{\includegraphics[width=3.5in]{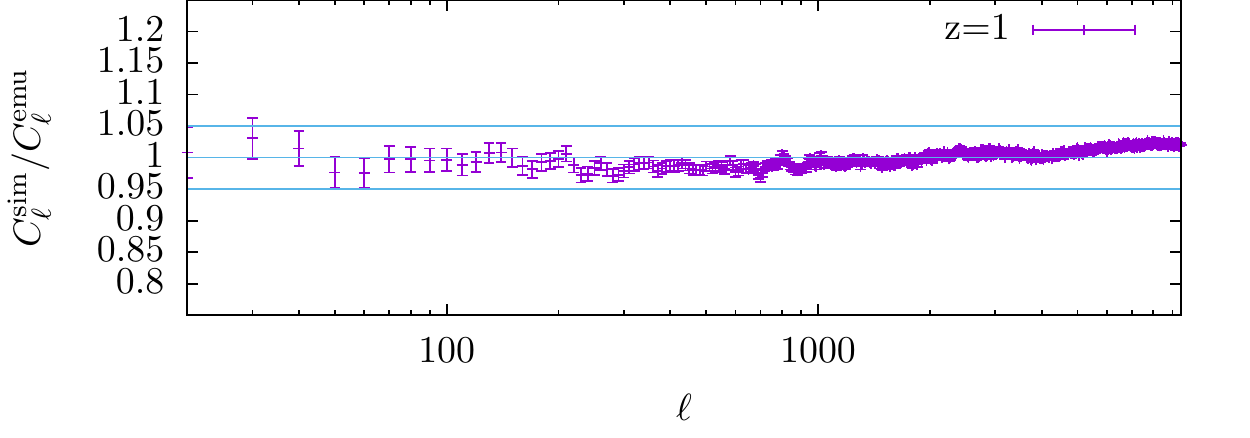}}
\caption{\label{fig:lightcone} 
Upper panel: Angular correlation function of particle overdensity in the light-cone outputs at $z = 0.5$ and $z = 1.0$ (see text for details). For comparison purposes, we show predictions computed from the Cosmic Emulator of~\cite{emu_ext}. We have chosen to plot every 10th multipole. 
The error bars show the 1-$\sigma$ predicted level of cosmic variance (neglecting non-Gaussian terms), corrected for the Polspice apodization kernel.}
Lower panels: Ratio of the measurements and emulator predictions at $z = 0.5$ and $z = 1.0$, with light blue bands showing a $5\%$ range. The results are consistent with the matter power spectrum shown in Fig~\ref{fig:pow}, and within the accuracy bounds reported in ~\cite{emu_ext}. 
\end{figure}

In order to verify the accuracy of our particle light-cone generation, we show the result for the measured angular correlation functions of two light-cone shells at redshifts $z = 1$ and $z = 0.5$. These are created using a nearest grid point density map on a HEALPix\footnote{{https://sourceforge.net/projects/healpix/}}~\citep{healpix} grid of Nside 8192, with the angular correlation function measured using Polspice\footnote{http://www2.iap.fr/users/hivon/software/PolSpice/} \citep{polspice} with a $15^{\circ}$ apodization window  (Figure~\ref{fig:lightcone}). These are compared to the angular correlation function predicted from the Cosmic Emulator power spectrum, taking into account both the finite shell width and the apodization kernel. The agreement over the range $20 < \ell < 10000$ is comparable to that of the matter power spectrum, with any light-cone extrapolation biases subdominant to the emulator errors.   

In angular space measurements, finite box lengths typically lead to errors at large scales and high redshifts, as displayed by, e.g., Fig.~1 of \citet{blaizot}. This is attributed to the presence of correlations due to repeated structures along the line of sight, as well as in the perpendicular direction for large angular scales. Given the large box size of the simulation, we choose not to apply box rotations in order to avoid discontinuities; as a result, a pencil beam along the axis will be replicated for the first time at $z \approx 1.8$ and the second time at $z \approx 12$, while very large-scale measurements will begin to experience replications at $z \approx 0.7$. Correspondingly, correlations measured at high redshifts on large scales require further characterization and are ultimately limited by the finite box length of the simulation.

\subsection{Halo Mass Function}
\label{sec:HMF}

\begin{figure}[b]
\centerline{
 \includegraphics[width=3.5in]{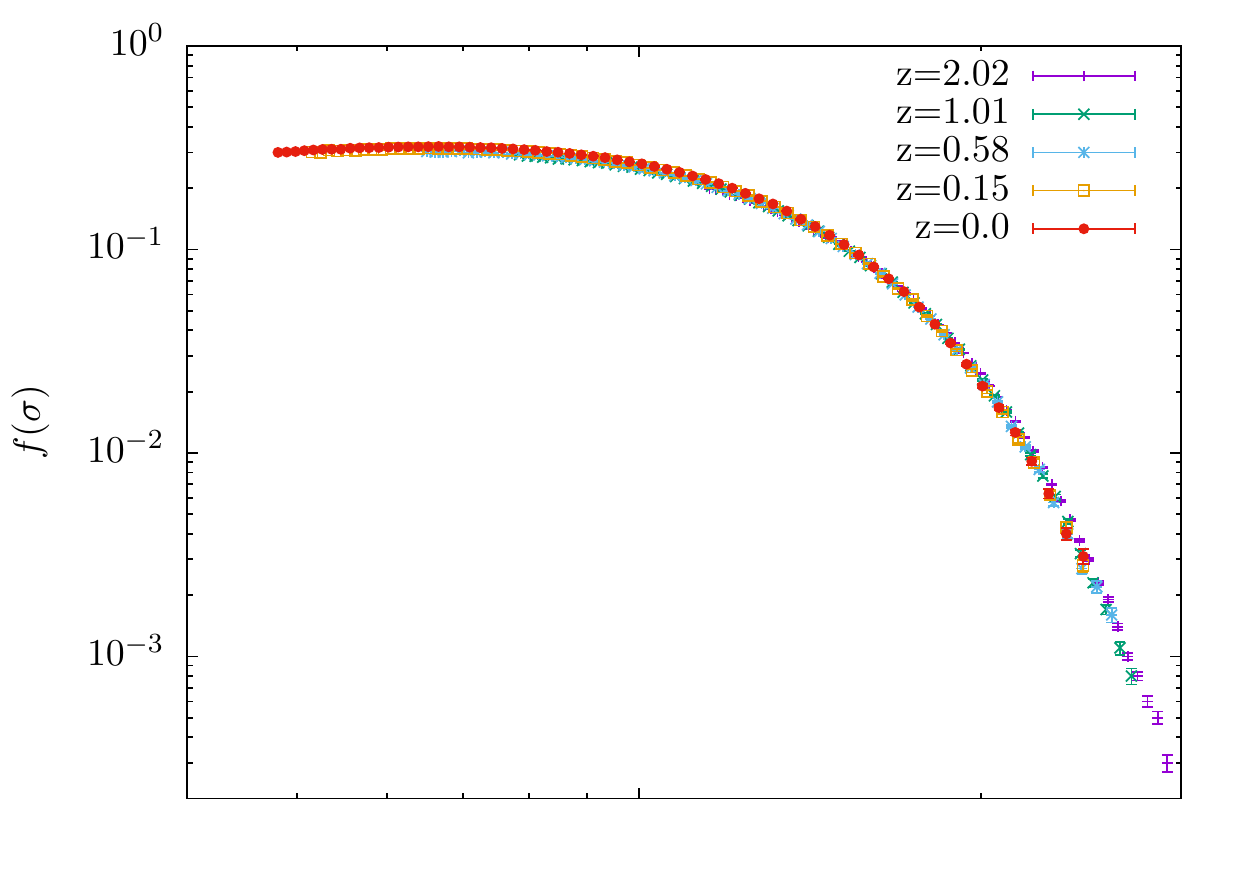}}
\vspace{-0.87cm}
\hspace{0.13cm}\centerline{\includegraphics[width=3.42in]{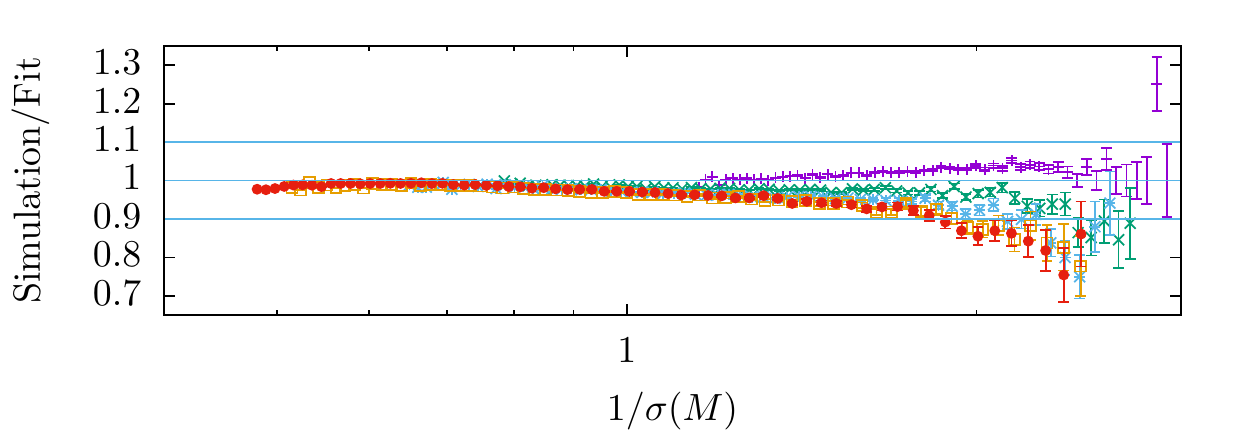}}
\caption{\label{fig:b02} Upper panel: Results for the differential mass function between redshifts $z=0$ and
  $z=2$ for FOF halos with a linking length of $b=0.2$. We show $f(\sigma,z)$ as a function of $1/\sigma$ to show the level of universality found as a function of redshift. Lower panel: Ratio of the simulation with respect to the mass function fit by~\cite{bhattacharya11}.
  The light blue lines show 10\% boundaries. Over a wide range, the agreement is well below 10\% discrepancy, at the higher mass, the agreement degrades slightly. This is in good agreement with previous studies of single large boxes from, e.g., the Outer Rim simulation~\citep{heitmann19b}.} 
\end{figure}

Next, we show measurements of the halo mass function. We focus on the FOF mass function obtained by using a linking length of $b=0.2$ and on predictions for overdensity masses $M_{200c}$. While we use a shorter linking length of $b=0.168$ for the main analysis of the simulations, the measurements employing the larger linking length have been widely studied in the literature. An important result due to \cite{J01} is that for this particular mass definition, the differential mass function $f(\sigma,z)$: 
\begin{equation}
f(\sigma,z)=\frac{d\rho/\rho_b}{d\ln
  \sigma^{-1}}=\frac{M}{\rho_b(z)}\frac{dn(M,z)}{d\ln[\sigma^{-1}(M,z)]}
\end{equation}
is close to universal. 
In this expression, $n(M,z)$ denotes the number density of halos with mass $M$,
$\rho_b(z)$ is the background density at redshift $z$, and $\sigma
(M,z)$ is the variance of the linear density field smoothed with a top-hat filter. Universality in this context means that a specific fitting function that encapsulates cosmology only through the linear power spectrum describes the mass function accurately across a range of redshifts. In this paper, we use the fitting function derived by \cite{bhattacharya11}:
\begin{equation}\label{fit1}
f^{\rm Bhatt}(\sigma,z)=A\sqrt{\frac{2}{\pi}}
\exp\left[-\frac{a\delta_c^2}
  {2\sigma^2}\right]\left[1+\left(\frac{\sigma^2}{a\delta_c^2}
  \right)^p\right] \left(\frac{\delta_c\sqrt{a}}{\sigma}\right)^q,
\end{equation}
with the parameters:
\begin{equation}\label{fit2}
A=\frac{0.333}{(1+z)^{0.11}},~~a=\frac{0.788}{(1+z)^{0.01}},~~p=0.807,~~q=1.795.
\end{equation}
The density threshold for spherical collapse, $\delta_c=1.686$, is
held fixed at all redshifts. In~\cite{QCont}, the fit included a small modification to simplify the redshift dependence -- $a$ was simply set to 0.788, eliminating the extra $z$-dependence. We compared our results to both fits and found that for $z=1$, the simplified version performed slightly better, while for $z=2$, the original form provided a closer fit.   

\begin{figure}[t]
\centerline{
 \includegraphics[width=3.5in]{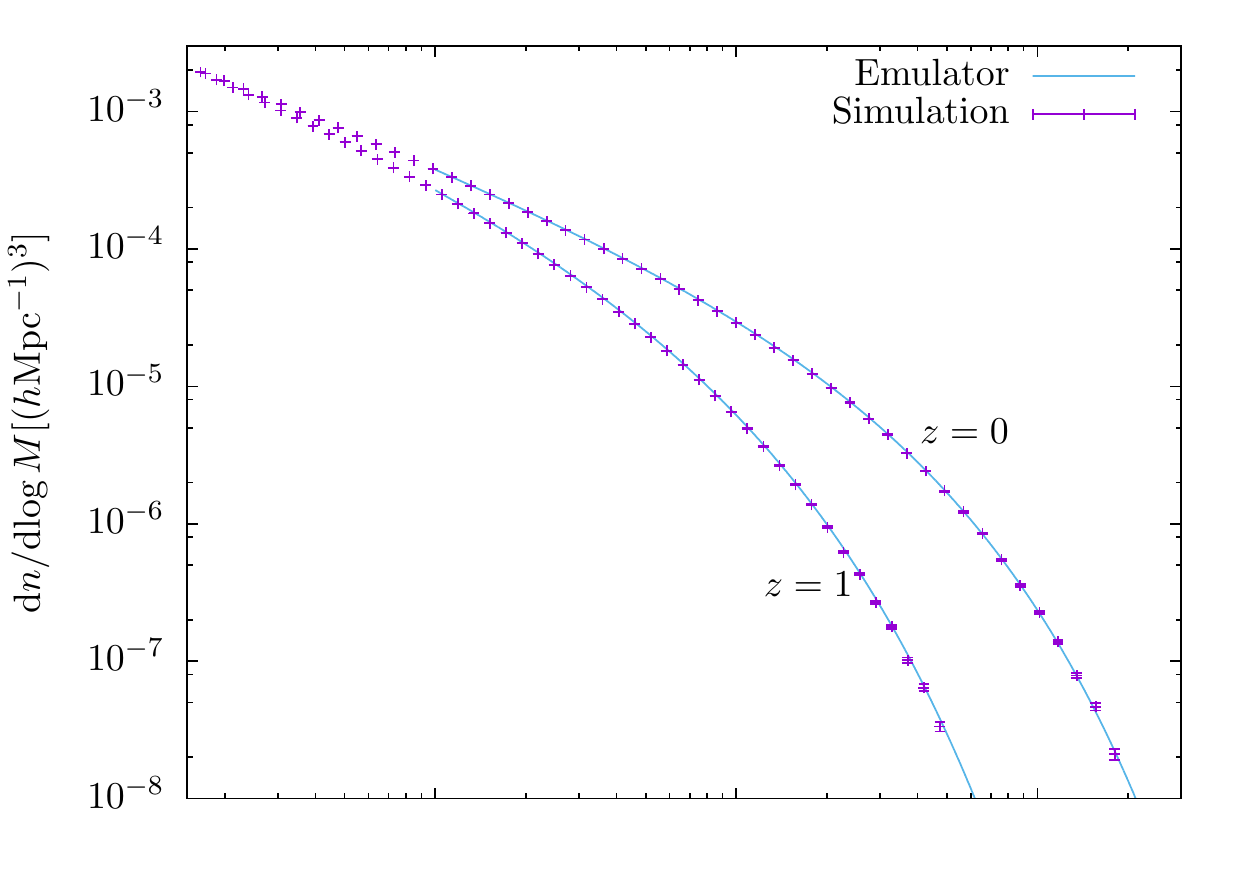}}
\vspace{-0.55cm}
\hspace{0.05cm}\centerline{\includegraphics[width=3.5in]{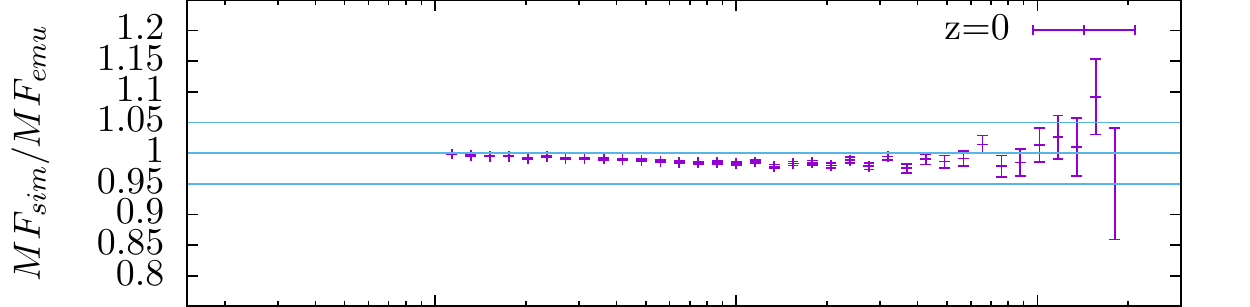}}
\vspace{-0.9cm}
\hspace{0.05cm}\centerline{\includegraphics[width=3.5in]{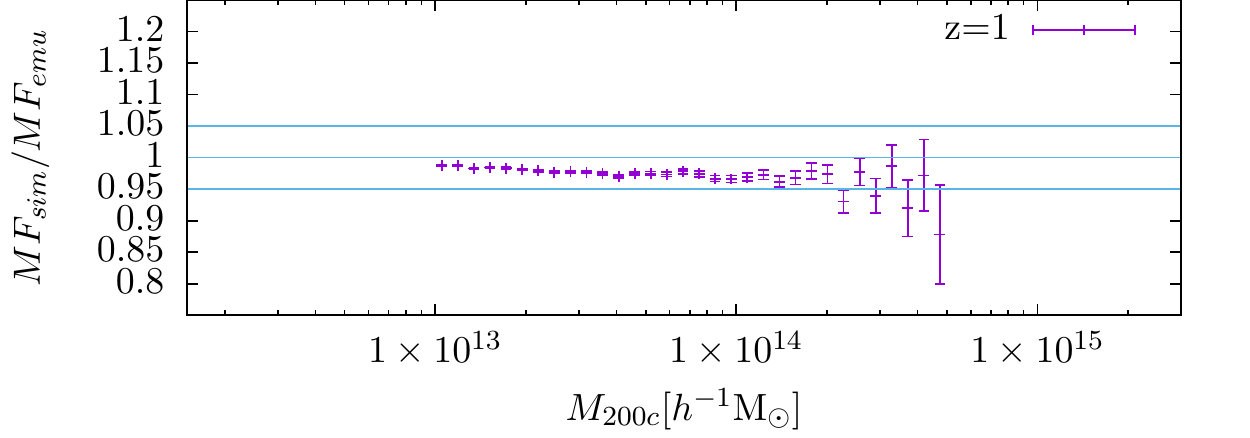}}
\vspace{0.5cm}
\caption{\label{fig:m200} Upper panel: Mass function results at redshift $z=0$ and
  $z=1$ for spherical overdensity halos, M$_{200c}$. For comparison, we show predictions from the Cosmic Emulator~\citep{emu_massf}. Lower panels: Ratio of the simulation and the emulator at $z=0$ and
  $z=1$. Note, that t}he emulator predictions start only at masses larger than $10^{13}h^{-1}$M$_\odot$ therefore reducing the mass range shown in the ratio panels. The light blue bands show a 5\% range. The results are within the accuracy bounds reported in~\cite{emu_massf}. 
\end{figure}

In order to account for FOF discreteness effects for low-mass halos, we apply the empirical correction given by
\begin{equation}
n_h^{\rm corr} = n_h(1-n_h^{-0.65}),
\end{equation}
where $n_h$ denotes the number of particles in a halo. A correction of this type was first introduced by~\cite{warren} and later recalibrated by~\cite{bhattacharya11}. We impose a halo mass cut and only include halos with masses larger than 10$^{12}h^{-1}$M$_\odot$ ($\sim$ 400 particles per halo). At this point, the correction is very small. In order to guarantee good sampling per mass function bin, we require each bin to have at least 100 halos. This restricts our measurement range slightly in the cluster regime. The error bars shown are Poisson error bars. As discussed in \cite{lukic07}, a slightly modified Poisson error given by $\sigma_\pm=\sqrt{n_h+1/4}\pm1/2$ introduced by \cite{Heinrich:2003} can be used if $n_h$ in some bins is very small. However, since we require at least 100 halos per bin, this modification is negligible, and the familiar form $\sigma_\pm=\sqrt{n_h}$ can be used to accurately determine the error bars. 

The measurements for the differential mass function $f(\sigma)$ are shown in Figure~\ref{fig:b02} for five redshifts between $z=2.02$ and $z=0$. The lower panel shows the ratio with respect to the fit given in Equation~(\ref{fit1}). The agreement is better than 5-10\% over most of the mass range, consistent with previous works. The discrepancy at higher masses is due the limited volume available from a single simulation. The fit derived in~\cite{bhattacharya11} is based on more than twice the volume available in the Last Journey simulation and is also based on a different cosmology. For a comprehensive discussion about the volume effects on the mass function, see, e.g.,~\cite{crocce10}.

Turning to the SO mass definition, Figure~\ref{fig:m200} shows a comparison between the simulation results at $z=0$ and $z=1$ with respect to a recently developed mass function emulator~\citep{emu_massf} based on the Mira-Titan Universe simulation suite, first described in~\cite{Heitmann:2015xma}. The Mira-Titan Universe includes a dynamical dark energy equation of state and neutrino mass in addition to the standard six cosmological parameters. The emulator's mass range is limited by the lower mass resolution of the Mira-Titan Universe simulations (around $m_p\sim 10^{10}$M$_\odot$, depending on the cosmology for each model) and provides predictions at the group and cluster scales, $m_{\rm halo}\ge 10^{13}h^{-1}$M$_\odot$. Overall, the agreement between the simulation and the emulator for both redshifts is around 2-3\%. As for the differential mass function, we employ Poisson error bars in Figure~\ref{fig:m200}. 

\subsection{Concentration-Mass Relation}

\begin{figure}[b]
\hspace{0.13cm}\centerline{\includegraphics[width=3.42in]{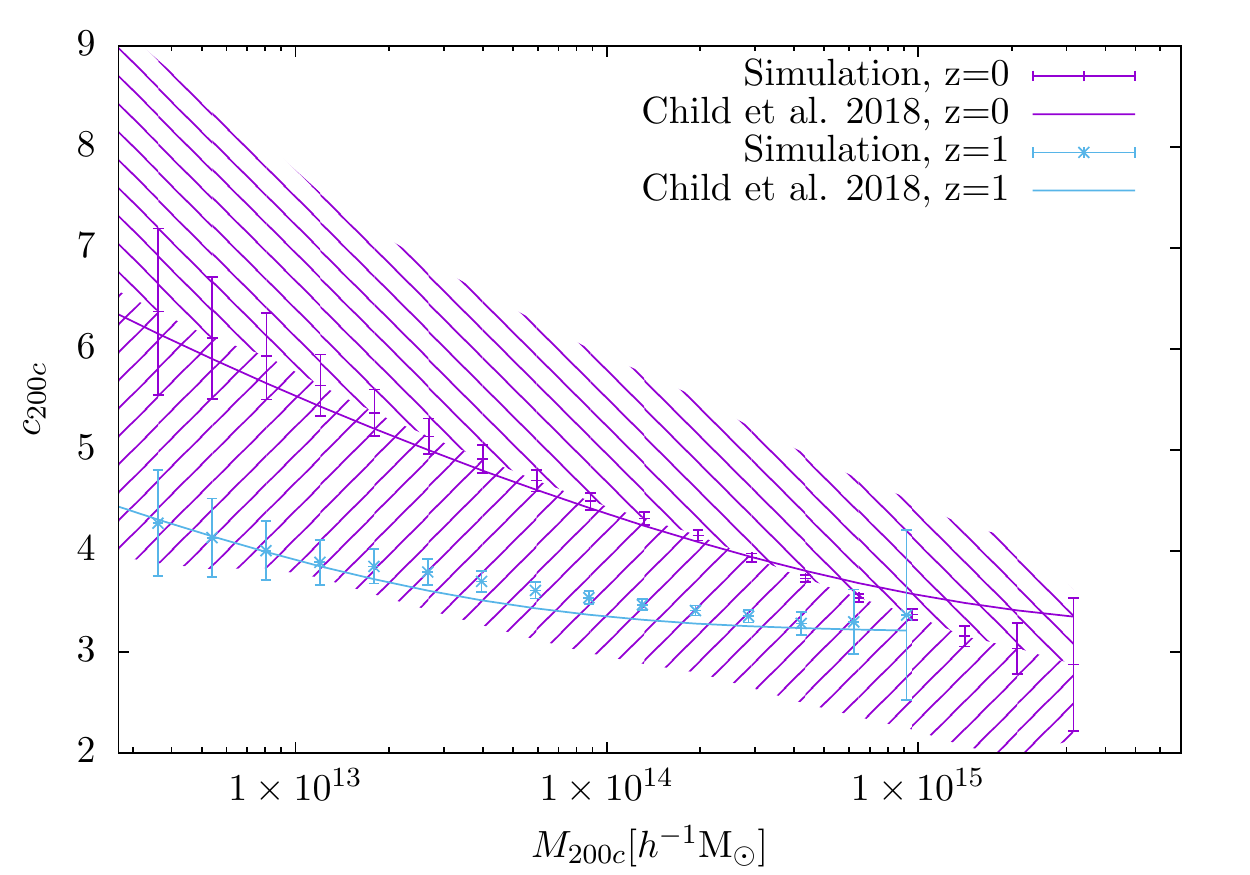}}
\caption{\label{fig:cm} Halo concentration-mass relation measurement for redshift $z=0$ (purple) and $z=1$ (light blue). For $z=0$, we show the 1-$\sigma$ standard deviation as the shaded region. In addition to the measurements from the simulation, we show the fitting function derived in \cite{child} (see Equation~\ref{fit}). } 
\end{figure}

We present results for the halo concentration-mass relation for two redshifts, $z=0$ and $z=1$. We measure the concentration by fitting a Navarro-Frenk-White (NFW) profile \citep{nfw1,nfw2} to each halo, as given by 
\begin{equation}
\rho(r)= \frac{\delta\rho_{\rm{crit}}}{(r/r_s)(1+r/r_s)^2},
\label{eq:nfw}
\end{equation}
where $\delta$ is a characteristic dimensionless density, and $r_s$ is
the scale radius of the NFW profile. The concentration of a halo is
defined as $c_{\Delta}=r_{\Delta}/r_s$, where $\Delta$ is the
overdensity with respect to the critical density of the
Universe, $\rho_{\rm{crit}}=3H^2/8\pi G$, and $r_{\Delta}$ is the
radius at which the enclosed mass, $M_{\Delta}$, equals the volume of
the sphere times the density, $\Delta \rho_{\rm{crit}}$. We compute
concentrations corresponding to $\Delta=200$, corresponding in turn to
$c_{200c}= R_{200c}/r_s$. Details about our fitting approach are given in 
\cite{bhattacharya13} and \cite{child}.  \cite{child} also discussed additional approaches for measuring halo concentrations and compared them. Following the results found in that paper, we use the profile fitting technique here, which provides robust measurements.  We compare our results to an approximate fit derived in \cite{child} (see their Equation 18 and Table 1):
\begin{equation}
\label{fit}
c_{200c}=A\left[\left(\frac{M_{200c}/M_\star}{b}\right)^m\left(1+\frac{M_{200c}/M_\star}{b}\right)^{-m}-1\right]+c_0,    
\end{equation}
with $A=3.44$, $m=-0.1$, and $c_0=3.19$ for $z=[0,4]$ and measurements for all (relaxed and unrelaxed) halos. This simple fit was developed to compare results from different cosmological simulations by making the assumption that most of the cosmology dependence can be encapsulated via the nonlinear mass scale, $M_\star$. Figure~\ref{fig:cm} shows our measurements and the fit for the c-M relation for $z=0$ and $z=1$; the result is given as a function of $M_{200c}$ and not of $M_{200c}/M_\star$ to enable easier comparison with the literature. (For the interested reader, comparisons of twelve different results from the literature with the above fit are given in \cite{child} in Figure~A14.) We use the publicly available {\sc Colossus} package developed by \cite{diemer} to determine the fit over the required mass ranges. Overall, the fit captures our results well for both redshifts. The error bars for bin $i$ are determined following the approach described in \cite{bhattacharya13}, Equation~7, by adding the error contribution from the individual concentration measurements and the Poisson error due to the finite number of halos in an individual bin in quadrature:
\begin{equation}
\Delta c_i (M)=\sqrt{\frac{\sum_j\delta c_j}{N_i}+\frac{c^2(M)}{N_i}},   
\end{equation}
where $\Delta c_i$ is the error for each bin and $\delta c_j$ the individual concentration measurement error for each halo. We have neglected the mass weighting that was applied in \cite{bhattacharya13}, Equation~7, after confirming that it only changes the result minimally.  As is apparent in Figure~\ref{fig:cm}, 
this error definition captures the larger uncertainty for the concentration measurement for small halos due to their less well-resolved profiles and for the larger halos due to their smaller number.

\subsection{Infall Merger Tree Masses}
\label{sec:infall}

\begin{figure}[b]
\centerline{\includegraphics[width=3.15in]{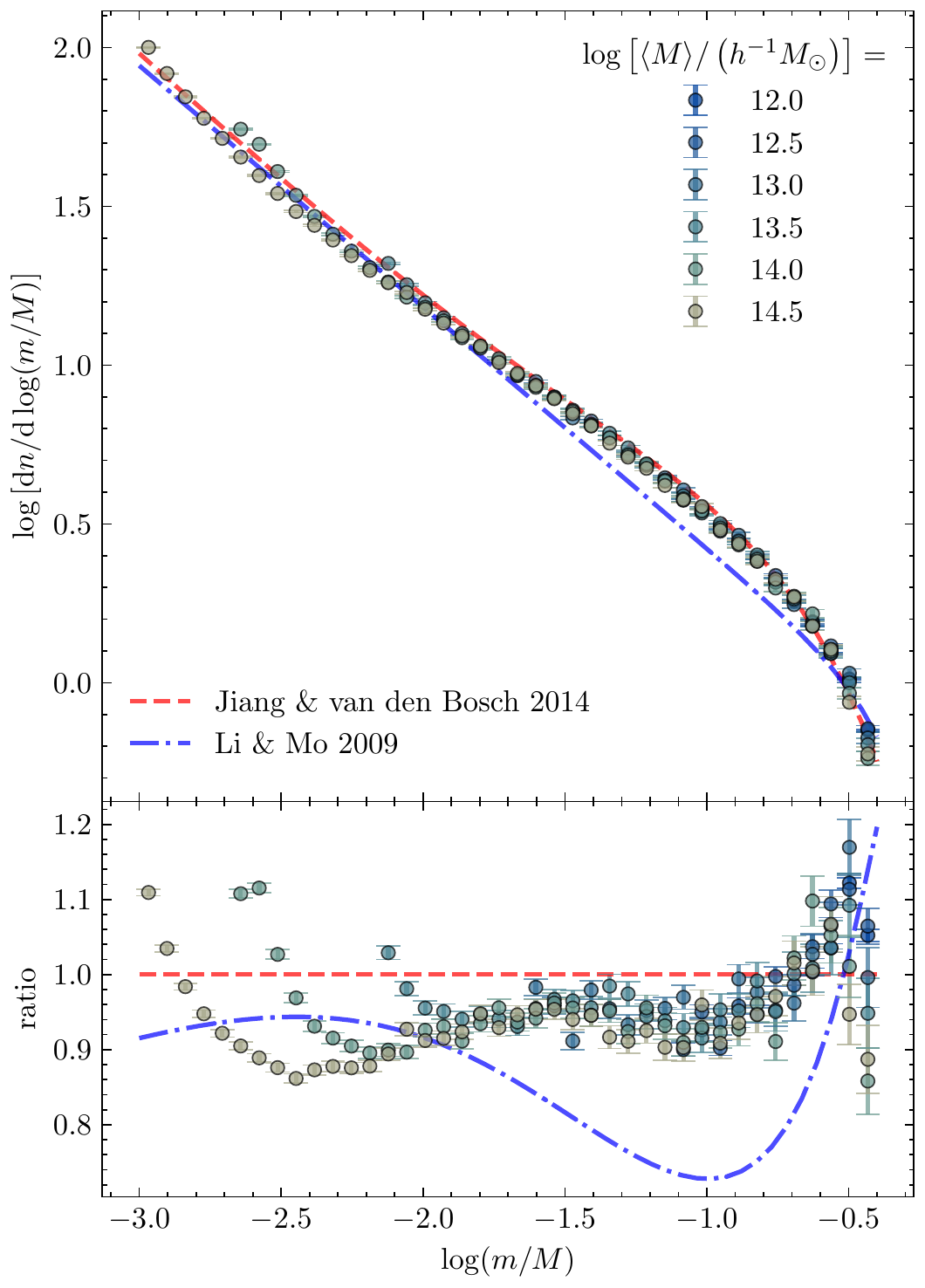}}
\caption{\label{fig:usmf_1storder} Upper panel: The infall halo mass function at redshift $z=0$ for first-order progenitors.
Our results (points) are the averages for the six host halo mass bins specified in Table \ref{tab:hostbins}.
An initial rise near the minimum halo mass of the simulation is not plotted for our host bins.
Host halo mass and infall mass are determined using the FOF masses recorded from the Last Journey simulation.
The fitting functions of \cite{jvdb14} (dashed line) and \cite{lm09} (dash-dotted line) for the unevolved subhalo mass function of first-order subhalos are also shown.
Both functions fit the Millennium Simulation results of \cite{lm09}, who have used a virial mass definition for the halos.
Lower panel: Ratios of our mass functions and the \cite{lm09} fit to the \cite{jvdb14} fitting function.} 
\end{figure}

\begin{table}[t] \caption{Host halo mass bins}
\begin{tabular}{cccc}
\hline\hline
$\log \left[ \frac{\langle M \rangle} { h^{{-1}}\mathrm{M_\odot} } \right]$ & $M_{min}$   & $M_{max}$   & Host halo \\
\hfill & [$10^{12}h^{{-1}}\mathrm{M_\odot}$]    &[$10^{12}h^{{-1}}\mathrm{M_\odot}$]   & count \\
\hline\hline
12.0 & 1.000   & 1.000   & 378,420          \\
12.5 & 3.163   & 3.163   & 42,640           \\
13.0 & 10.000  & 10.005  & 14,303           \\
13.5 & 31.625  & 31.679  & 10,102           \\
14.0 & 100.000 & 100.674 & 10,018           \\
14.5 & 316.229 & 328.172 & 10,001    \\
\hline\hline
\end{tabular}\label{tab:hostbins}
\end{table}

In this section, we show a subset of the results from our merger trees by studying the infall halo mass functions (often referred to as ``unevolved subhalo mass functions'') and comparing to previous results. The infall halo mass function describes the mass distribution of halos just prior to merging with a more massive host. More information about merger tree statistics will be reported elsewhere.

We first calculate the first-order infall halo mass function (as defined in \cite{jvdb14}) of our halo merger trees corresponding to measuring the infall masses of all halos that were accreted 
along the most massive progenitor branch of a given merger tree. The infall halo mass function $\text{d}n/\text{d}\log(m/M)$ describes halos with infall mass $m$ that were accreted onto any given tree with a final halo descendant mass $M$ at $z=0$. (All halo masses here are FOF masses with linking length $b=0.168$ and the reported error bars are Poisson errors measured in each bin.) Note that these substructures are obtained from the core catalog using the host core tag described in Section~\ref{sec:CCG}, where only cores with host core tags equal to the final central tag at $z=0$ were accreted by the main progenitor halos. The upper panel of Figure~\ref{fig:usmf_1storder} shows the infall mass function for six $z=0$ host halo mass bins. Table \ref{tab:hostbins} lists the average host halo mass, mass range, and number of host halos in each bin.

\begin{figure}[t]
\centerline{\includegraphics[width=3.15in]{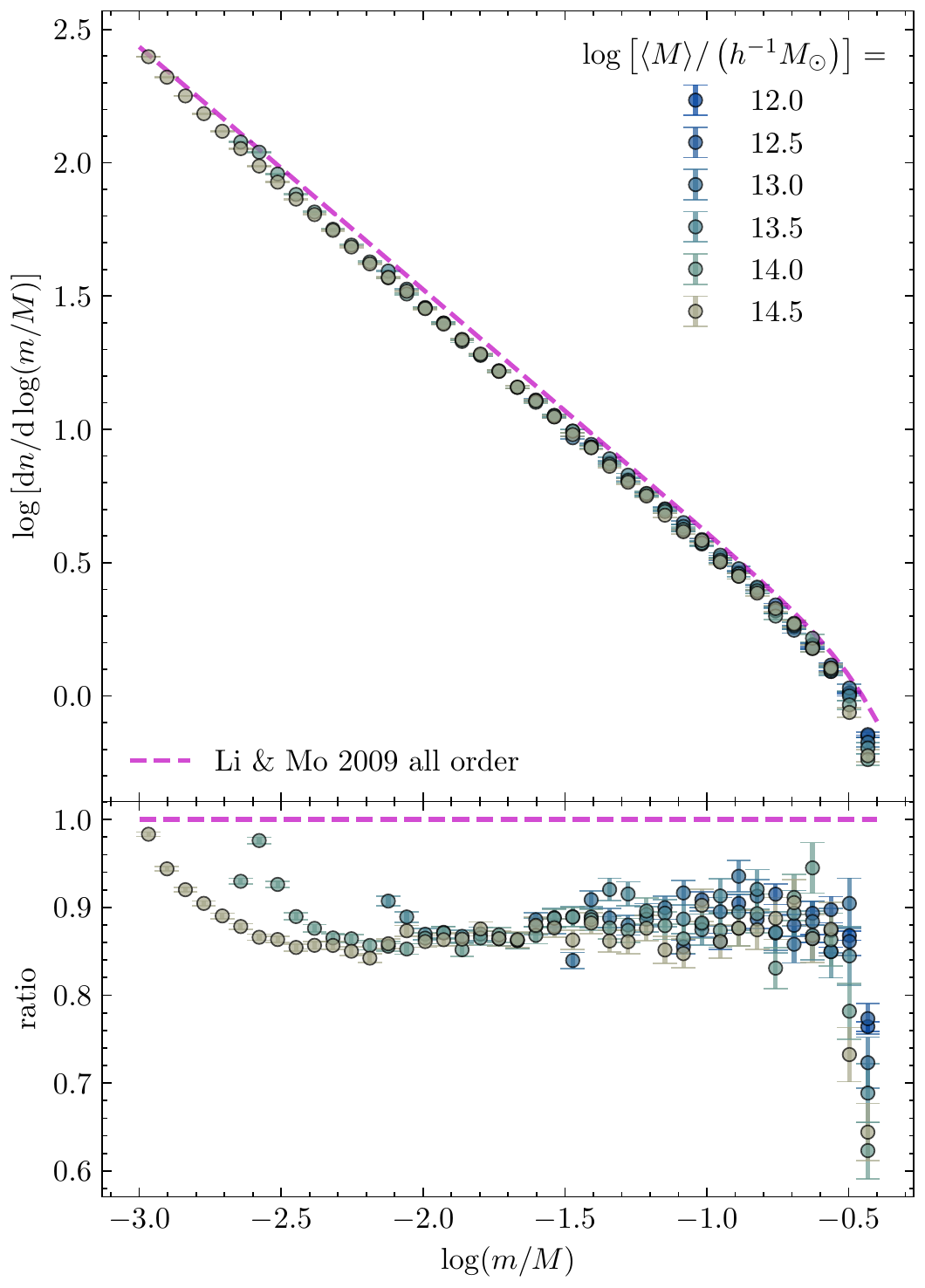}}
\caption{\label{fig:usmf_allorder} Upper panel: The infall halo mass function at redshift $z=0$ for all-order progenitors. Our results (points) are the averages for the six host halo mass bins specified in Table \ref{tab:hostbins}.
An initial rise near the minimum halo mass of the simulation is not plotted for our host bins. The fitting function of \cite{lm09} for the unevolved subhalo mass function of all-order subhalos (dashed line) is also shown. As in Figure~\ref{fig:usmf_1storder}, the Last Journey results employ FOF masses, while the fitting function is based on virial masses. Lower panel: Ratio of our mass functions to the \cite{lm09} fitting function. Note that the fit was derived for a cosmology with higher value for $\sigma_8=0.9$ and a different mass definition for the host halo, likely accounting for the difference in the amplitude.} 
\end{figure}

We include a comparison with fitting functions derived in \cite{jvdb14} and \cite{lm09} for the unevolved subhalo mass function of first-order subhalos in the Millennium Simulation.
\cite{jvdb14} noted that past work, including \cite{lm09}, have found the unevolved subhalo mass function of first-order subhalos to be approximately independent of host halo mass for a $\Lambda$CDM cosmology. 
This is consistent with the lack of mass dependence in our results, as shown in Figure~\ref{fig:usmf_1storder}.
Our results are in qualitative agreement, lying in between the two fits. (We note that comparison is not fully direct because the mass ratios for the Last Journey results are for FOF masses, while the analytic fits use virial mass ratios.)

In addition, we calculate the ``all-order'' infall halo mass function at $z=0$ of our halo merger trees.
This is the mass function of all halos that have merged into either a $z=0$ host halo or any of its progenitors.
The upper panel of Figure \ref{fig:usmf_allorder} shows the infall mass function for the same six host halo mass bins used in Figure~\ref{fig:usmf_1storder} (see Table \ref{tab:hostbins}).
We include a comparison with the fitting function of \cite{lm09} for the unevolved subhalo mass function of all-order subhalos in the Millennium Simulation. The small difference in the amplitude is most likely due to the different host halo mass definition used in \cite{lm09} and the different cosmology employed. A more detailed investigation is forthcoming (Sultan et al., in preparation).  

\section{Data Release}
\label{sec:release}

\begin{table*}\caption{Released data products}
\begin{center}
\begin{tabular}{ccc}
\hline\hline
     Data Product & File Content & Overall Size \\
     \hline\hline
     Downsampled particles, 1\%& $(x,y,z)$, $(v_x,v_y,v_z)$, particle tag& 3TB \\
     Halo particles, 1\%, minimum 5 per halo & $(x,y,z)$, $(v_x,v_y,v_z)$, particle tag, halo tag& 2.4TB\\
     FOF properties & halo tag, m$_{\rm halo}$, $(x,y,z)_{\rm pot}$, $(x,y,z)_{\rm COM}$, $(v_x,v_y,v_z)_{\rm COM}$ & 622GB\\
     \hline\hline
\end{tabular}\label{tab:release}
\begin{tablenotes}
          \item 
          \small Note: All data products are released at the following redshifts: $z=\{1.49, 1.43, 0.86, 0.78, 0.54, 0.50, 0.21, 0.05, 0.0\}$.
        \end{tablenotes}
\end{center}        
\end{table*}

As part of this paper, we make a subset of the simulation data from the Last Journey publicly available.  We use the web-based data portal introduced in \cite{heitmann19a} for this purpose\footnote{https://cosmology.alcf.anl.gov}. The data portal is based on Petrel\footnote{https://press3.mcs.anl.gov/petrel}, a pilot project for data management and sharing, hosted at the ALCF. The web portal allows easy browsing of the available data products. Once the data sets of interest have been identified, Globus\footnote{https://www.globus.org} enables fast and secure data transfer from Petrel to the target location. The data are stored in a customized HACC format, genericIO. This format was developed as an optimized read-and-write functionality for HACC at scale. The data portal provides a link to the genericIO repository\footnote{https://xgitlab.cels.anl.gov/hacc/genericio.git}, which includes an example Python module for accessing the data. 

The size of the full data set (close to one PB) by far exceeds the available storage for our project on Petrel; therefore, we carefully chose a subset of the data for the public release. The data products we make available -- halo catalogs, halo particles, and a subset of the full particle dataset at nine redshifts -- enable the creation of synthetic sky catalogs targeted at cosmological survey science. For example, very similar data products from the Outer Rim simulation were used by the eBOSS collaboration to build HOD-based catalogs to enable clustering measurement of quasars via the correlation function~\citep{eboss1} and the power spectrum~\citep{eboss2}, and by the anisotropic clustering of quasars in configuration space~\citep{eboss3}. The downsampled particles allow the measurement of correlation functions and the generation of density and potential fields for a range of studies.

More specifically, we provide data products for the following nine redshifts:
\begin{equation}
z=\{1.49, 1.43, 0.86, 0.78, 0.54, 0.50, 0.21, 0.05, 0.0\}.
\end{equation}
We enable access to the downsampled particle snapshots that contain 1\% of the full particle outputs, randomly selected. In addition, we provide the results from the halo finder, including a range of halo properties. We release results from the FOF, $b=0.168$ finder at the same nine redshifts listed above. In particular, we provide a halo tag (which is reassigned at each snapshot and hence cannot be used to track halos over time), $m_{\rm halo}$ [$h^{-1}$M$_\odot$], a halo center ($x,y,z$) [$h^{-1}$Mpc], measured at the halo's potential minimum, a halo center of mass ($x,y,z$) [$h^{-1}$Mpc], and a mean velocity of the particles within the halo ($v_x,v_y,v_z$) [km/s]. The third data product we release are 1\% of the particles that reside in halos. For all halos with less than 500 particles, we provide information for 5 halo particles. For each particle, we provide its position and velocity, a particle tag (which is the same at each snapshot), and the halo tag to connect the particles back to their host halo. All velocities for particles and halos are measured in comoving peculiar units. We summarize the released data products in Table~\ref{tab:release}.

\section{Summary and Outlook}
\label{sec:summary}

Large-scale simulations have started to play a very important role in survey science over the last few years. Supercomputers have reached the level where simulations can cover enough volume and have sufficient resolution to enable the creation of realistic sky maps that can be used by cosmological surveys to investigate the correctness of their pipelines; model, investigate and understand systematic effects; and carry out meaningful blind data challenges. 

In this paper, we have described the Last Journey simulation, an extreme-scale gravity-only run, carried out on the Mira supercomputer. We evolved more than 1.2 trillion particles in a (3.4$h^{-1}$Mpc)$^3$ volume using a cosmological model with parameter choices consistent with the analysis reported by the~\cite{planck18}. We presented results for standard measurements of the simulation, including power spectra and mass functions, and compared them against results from emulators, finding very good agreement. 

As cosmological simulations continue to grow in size, resource management for data analysis has become a key concern. There is increasing emphasis on carrying out in situ analyses as the code is running, with the aim of minimizing I/O and offline computing and storage. The Last Journey run provides a good example of this imperative, as cosmological simulations prepare for the coming exascale era.

This paper describes the data sets available from the Last Journey run. These include particle light cones and detailed halo properties starting at a redshift $z\approx 10$. Merger trees and core catalogs have also been produced. The use of core tracking integrated with merger trees provides a new method to follow the evolution of substructures within halos in very large simulations. Selected outputs from the simulations are publicly available at the HACC Simulation Data Portal.\footnote{https://cosmology.alcf.anl.gov}

The simulation is designed to support a wide range of cosmological surveys across different wavebands. An important next step is the creation of a synthetic galaxy catalog using a semi-analytic approach combined with empirical relationships to provide good fits to a host of observations. The core catalog will enable the tracking of substructures, serving as a replacement for the commonly used subhalo merger tree and allowing for more accurate placements of model galaxies within the halos in the simulations. The core-tracking approach in combination with the availability of particle light cones also facilitates accurate small-scale lensing measurements, while the presence of tidal field and halo shape information will allow incorporation of intrinsic alignment modeling into such catalogs, adding realistic complexity and enabling testing of mitigation techniques. The presence of halo light-cone outputs out to high redshifts of $z \approx 10$ will enable improved modeling of the cosmic infrared background (CIB) in support of next-generation CMB observations. The simulation outputs are also well suited for the creation of  galaxy-galaxy and cluster weak lensing measurements, given the availability of full particle light cones of $\sim10^9$ mass resolution at $z<3$.  Additional outputs of downsampled particle light cones at higher redshifts ($3<z<10$) will be used for creating CMB lensing maps. 

The current paper is the first in a series of at least three Last Journey papers. The second paper will provide a detailed description of the core-tracking approach and the connected mass modeling method that is essential to enable the substitution of subhalo merger trees with core catalogs. The third paper will describe the results obtained from a SAM approach augmented with empirical modeling to build a detailed synthetic sky map.

\begin{acknowledgments}

We thank Malin Renneby for many early discussions regarding the project. We are grateful to Michael B\"{u}hlmann and Claude-Andr\'e Faucher-Gigu\`ere for a careful reading of the manuscript and for valuable comments. Argonne National Laboratory's work was supported under the
U.S. Department of Energy contract DE-AC02-06CH11357.  An award of computer time was provided by the ASCR Leadership Computing Challenge (ALCC) program. This research was supported in part by DOE HEP's Computational HEP program. This
research used resources of the Argonne Leadership Computing Facility
at the Argonne National Laboratory, which is supported by the Office
of Science of the U.S. Department of Energy under Contract
No. DE-AC02-06CH11357. We are indebted to the ALCF team for their
outstanding support and help in enabling us to carry out a full-scale simulation on Mira. 

\end{acknowledgments}

\end{document}